\documentclass[12pt,nofootinbib,longbibliography]{revtex4-1}
\usepackage{amsmath,amssymb,ytableau, graphicx}

\begin{document}

\title{The endpoint of partial deconfinement}
\author{David Berenstein $^\dagger$, Kai Yan $^{\dagger, \ddagger}$ }
\address{ $^\dagger$ Department of Physics, University of California, Santa Barbara, CA 93106}
\address{$^\ddagger$ Department of Physics,
The University of Chicago, 933 East 56th Street, Chicago, Illinois 60637, USA}

\begin{abstract}
We study the matrix quantum mechanics of two free hermitian $N\times N$ matrices subject to a singlet constraint in the microcanonical ensemble. This is the simplest example of a theory that at large $N$ has a confinement/deconfinement transition. In the microcanonical ensemble, it also exhibits partial  confinement with a Hagedorn density of states. We argue that the entropy of these configurations, calculated by a counting of states based on the fact that Young diagrams are dominated by Young diagrams that have the VKLS shape. When the shape gets to the maximal depth allowed for a Young diagram of $SU(N)$, namely $N$, we argue that the system stops exhibiting the Hagedorn behavior. The number of boxes (energy) at the transition is $N^2/4$, independent of the charge of the state.

\end{abstract}

\maketitle
\section{Introduction}

The confinement/deconfinement transition plays an important role in the study of gauge theories. 
Thanks to the AdS/CFT correspondence, the confinement/deconfined phase can be associated to spacetimes with and without a black hole \cite{Witten:1998zw}. In the gravity side, this transition is the Hawking-Page first order phase transition \cite{Hawking:1982dh}. 
The physics in AdS tells an additional story. For low energies, there is a Hagedorn density of states (basically, we have a spectrum of strings propagating in an AdS spacetime). 
The Hagedorn temperature and details of the phase transition were studied perturbatively in \cite{Sundborg:1999ue,Aharony:2003sx}. The Hagedorn temperature in ${\cal N}=4 $ SYM at large $N$ has been computed more recently using methods of 
integrability \cite{Harmark:2017yrv,Harmark:2018red,Harmark:2021qma,Ekhammar:2023glu}.
Currently, this part of the behavior of the duality at low energies with respect to $N^2$, but still large energies compared to the string scale can be claimed to be well understood.

Usually, in the study of first order phase transitions, there is a Maxwell construction that lets one fix the temperature at the transition temperature and one can vary the energy by occupying different regions of space with different phases of the theory. This is a coexistence between two phases. This way, the temperature stays fixed when one varies the energy. In the Hagedorn setup, the exponential growth of states fixes the temperature by different means and usually occurs at a higher temperature than the first order Hawking-Page phase transition. However, as shown in \cite{Aharony:2003sx} (see also \cite{Sundborg:1999ue}), at zero coupling
the two transitions are the same. 
From the point of view of black hole physics, small black holes have negative specific heat, while large  black holes have positive specific heat. The small black holes are thermodynamically unstable in the canonical ensemble. If one fixes the energy instead of the temperature, one can have negative specific heat. This just indicates a faster growth of entropy with the energy than one would naively imagine. Basically, one needs $(\partial^2_ES)>0 $ to get a negative specific heat. The Hagedorn behavior $S\propto E$ sits exactly at infinite specific heat, and any perturbation can in principle turn the specific heat negative.

These arguments suggest that there should be a notion of a Maxwell construction between 
two phases that describes the Hagedorn behavior at zero coupling, as the Hagedorn and the 
confinement/deconfinement transition coincide. The thermodynamic limit in this setup is associated with phase transitions at large $N$, so the growth of states is produced by growing the size of the gauge group, not the volume of space. 
A notion of a mixture of confinement and deconfinement should occur in the variables that are becoming a thermodynamic volume. In this case, the notion of volume is in the labels of the internal degrees of freedom of the matrices themselves. This idea was proposed as a way to understand the small black holes in AdS space \cite{Asplund:2008xd,Jokela:2015sza,Hanada:2016pwv}.

A notion of a subgroup being deconfined, while the rest is confined is called  partial deconfinement (see \cite{Hanada:2022wcq} 
for a short review). A natural question is if the process of going from partial deconfinement to full deconfinement is a crossover, or if there is a phase transition that separates them.
in \cite{Hanada:2018zxn} it was argued that there is a phase transition closely related to the Gross-Witten-Wadia \cite{Gross:1980he,Wadia:1980cp} transition separating partial deconfinement and confinement. Similar observations about phase transitions at large $N$ related to deconfinement are found in \cite{Dumitru:2004gd,Nishimura:2017crr} (see also \cite{Asano:2020yry}).

The main issue to be concerned about is that if one wants to understand the phase transition well, one needs to fix the energy, rather than the temperature.
Standard path integral methods in imaginary time work well if one fixes the temperature. Fixing the energy is not as simple. Counting states directly can be very hard. This is why it is important to have simple models where the behavior one wants to study can be understood in detail.

In this paper, we study such a simple model. The model we consider is the theory of two free $N\times N$ hermitian matrix quantum mechanics, subject to a singlet constraint. For simplicity, the angular frequency of the matrices is set equal to one, so that the energy and the occupation number are the same.  This gauge theory is one of the simplest that exhibits Hagedorn behavior and where partial deconfinement has been argued to be valid \cite{Berenstein:2018lrm}. It has also been argued that generic corrections can turn the specific heat negative \cite{Berenstein:2018hpl} as would be expected from a system that could in principle  describe small AdS black holes. The theory also has a conserved charge, so one can study the model as a function of both energy and charge.
Large $N$ counting suggests that we parametrize the information in terms of $\epsilon=E/N^2$, and the fraction of charge to energy $q=Q/E$. The large $N$ transition is studied by taking $N\to\infty$ keeping these quantities fixed. 

In this short note, we study the counting of states combinatorially using techniques from representation theory and tensor products of said representations. 
The goal is to better understand what fraction of the gauge group is partially deconfined as a function of the energy/charge and to use this information to make predictions about the locus of the phase transition. The states are determined by triples of Young diagrams and their degeneracy within this representation is given by squares of Littlewood Richardson coefficients. 
The number of boxes in the young diagrams is the total occupation number of each of the matrices and the total occupation number. 
The large $N$ limit requires large representations and a lot of our results are related to the asymptotic growth of Littlewood-Richardson  coefficients, following results in \cite{PAK201944}. The most important information is the typical shape of the Young diagrams that realize these estimates is the VKLS shape, attributed to Vershik, Kerov, Logan, and Shepp \cite{vershik1985asymptotic,logan1977variational} and that these dominate the entropy.
The transition occurs when the typical shape, rescaled to the number of boxes, reaches the maximum depth allowed by $SU(N)$ representations. This occurs at an energy $E=N^2/4$ (there are subleading corrections in $N$) regardless of the charge of the state. 

The paper is organized as follows. In section \ref{sec:count}, we introduce the model we study and the method of counting states using representation theory and young diagrams. We explain that the counting of states is computed by adding squares of  Littlewood-Richardson coefficients and that these must become large. Basically, there are too few Young diagrams to give the correct counting of states, so the counting is mostly on the multiplicity of the representation counting. We then use results in combinatorics to show that at large energy, the state counting is dominated by a specific shape: the VKLS shape. If a transition from partial deconfinement to deconfinement is to occur in the microcanonical ensemble, we conjecture that this must happen when the VKLS shape becomes disallowed at finite $N$ (the shape has the depth that is greater than $N$ as a function of the number of boxes in the Young diagram),
This predicts a specific energy for the transition at large $N$.

In section \ref{sec:numeric} we address our conjecture numerically. We do this by computing the degeneracy numerically: we compute the Littlewood Richardson coefficients and verify that the shapes that maximize these are VKLS: they are close to minimizing the hook length for a fixed number of boxes. We also observe that there are non-trivial critical exponents once the energy gets larger than the transition energy, verifying that the transition is weakly first order (it is somewhere between first and second order). We present numerical evidence for the main claim: that  the transition happens at a fixed energy $E=N^2/4$ regardless of the charge of the state.

Finally, in section \ref{sec:con} we conclude.

\section{Counting states and the typical Young tableaux} \label{sec:count}

The system we will be studying is a matrix quantum mechanics of two hermitian matrices $X,Y$. The Hamiltonian is given by
\begin{equation}
    \hat H= \frac12 \hbox{tr} \left[ P_X^2+P_Y^2+  X^2+ Y^2\right]
\end{equation}
and notice that each of the $2N^2$ oscillators has angular frequency $\omega=1$. Since the theory is free, the energy becomes identical as the occupation number plus the zero point energy. For convenience, we set the energy of the ground state to zero.  The occupation number of $X$ and $Y$ are also conserved separately, we can call the difference of these occupation numbers the charge of the state. The system also has a $O(2)$ symmetry that rotates $X$ into $Y$ which is a different symmetry and we will not be concerned with it directly.

The system has a $SU(N)$ symmetry that acts by conjugation $X\to U X U^{-1}, Y\to U Y U^{-1}$.
We will restrict to the singlet state under the $SU(N)$ symmetry. Our goal is to analyze this system in the microcanonical ensemble at large $N$ and at different values of the energy.
The scaling must be such that $E= N^2 {\cal \epsilon}$, where $\epsilon$ is a normalized energy divided by the growth of states at large $N$. We will do similar rescalings with the entropy.

The first goal is to show the Hagedorn behavior of states for this system. There is a representation of the counting of states in terms of traces. However, it is more instructive to start with the partition function at infinite $N$ for the singlet states. This has been computed in \cite{Sundborg:1999ue,Aharony:2003sx}. The generating function of states is given by 
\begin{equation}
Z=\prod_{n=1}^\infty\frac1{1 -x^n-y^n}\label{eq:part}
\end{equation}
The $x$ powers count how many $X$ are excited and the $y$ powers count how many $Y$ are excited in total and should be assumed to be positive real variables.

The partition function is convergent so long as $x+y<1$. Let us concentrate on the first term
\begin{equation}
    Z_1= \frac 1{1-x-y}= \sum_{k=0}^{\infty} (x+y)^k  
\end{equation}
If we fix the energy $E$, we need to fix $k=E$. There are exactly $2^k$ states accounted for in this sum (fix $k$ first and set $x=y=1$). These are all the possible words made of $x,y$ that have a length of exactly $k$. Each letter can be chosen to be $x$ or $y$ at any position of the word. 

If we include the other terms from the product, we have additional positive contributions.
This shows that the number of states  grows at least exponentially with the energy, giving us a Hagedorn behavior. The entropy would be $S\geq k \log 2 =  E\log 2$.  
Using thermodynamic relations $T dS = dE$, we find a temperature $T= (\log(2))^{-1}= \beta^{-1}$

We can now also fix the charge $Q=(n_x-n_y)/2$.
When we expand the term $(x+y)^k$, we get
\begin{equation}
(x+y)^k = \sum_{n_1+n_2=k} {k \choose n_1} x^{n_1} y^{n_2}
\end{equation}
This can also be interpreted probabilistically. The probability of getting an $X$ is $n_1/k= (1/2 +q) $ and the probability of getting a $y$ is $n_2/k= (1/2-q)$, where we have introduced the average charge per unit letter
\begin{equation}
    q= \frac{ Q} {E} = \frac12 \left(\frac { n_1-n_2 }{n_1+n_2}\right)
\end{equation}
and $-\frac12 \leq q\leq\frac 12$.
The (Shannon) entropy of such words is the number of letters times the entropy per letter
\begin{equation}
S= E(- p_x \log(p_x) -p_y \log (p_y)) 
\end{equation}
The expression $\beta_q =- p_x \log(p_x) -p_y \log (p_y) $ can be interpreted as an effective inverse temperature. In terms of $q$, it is given by
\begin{equation}
    \beta_q= -\left(\frac 12 +q\right) \log\left(\frac 12 +q\right) - \left(\frac 12 -q\right) \log\left(\frac 12 -q\right),
\end{equation}
and we notice that when we set $q=0$ we recover the original result for arbitrary words  $\beta= \log(2)$.

Now, let us consider finite $N$ at large temperature. In this limit, a classical physics computation  should be accurate. We have $2N^2$ degrees of freedom and we have $N^2$ constraints.  It is easy to argue, by a scaling argument (see \cite{Asplund:2012tg} for example)  that one should have exactly $E= N^2 T$ in this classical limit.
The basic idea is that the Gibbs partition function is given by
\begin{equation}
    Z\sim \int (d^{2N^2 }p) (d^{2N^2 }q) \delta^{N^2}( p\cdot q) \exp(-\beta p^2/2-\beta q^2/2)
\end{equation}
and one can then rescale $p,q$ to eliminate $\beta$ from the exponent. The quadratic constraints of the gauge transformations are schematically written as $p\cdot q$, understanding that these are $N\times N$ matrices of constraints.
The measure scales like $\beta^{-2N^2}$ and each delta function constraint, which is quadratic in $p,q$, scales like $\beta$, giving a total of $\beta^{N^2}$ from the constraints.  This leaves us with a total scaling of $\beta^{-N^2}$. This is the same as a partition function with $N^2$ harmonic oscillator degrees of freedom in phase space. If one adds a delta function of the energy, one needs to replace $N^2$ by $N^2-1$ above. 

The entropy, by the thermodynamic relation $T dS= dE$ then behaves as $S\simeq N^2 \log(T) \sim N^2 \log (E/N^2)$. In this regime, the entropy only grows logarithmically with the energy as opposed to linearly in the energy.
Notice that we are not taking into account the integration constant for the entropy carefully, as is standard in a classical calculation. 

Since the behavior at low and high energies are very different, there must be either a crossover from the Hagedorn behavior described above $S\propto E$, or an actual phase transition at large $N$ that separates these two behaviors. 
Both of these possibilities,  by large $N$ counting, should occur at an energy that scales with $N^2$. The idea of partial deconfinement versus full deconfinement is that this change of behavior is actually a continuous phase transition: that some quantities become discontinuous at some value of the energy with some non-trivial critical exponents.  One needs to keep $\epsilon=E/N^2$  finite when taking $N\to \infty$ to see the phase transition.
We're being very careful here to state that the transition occurs at a fixed energy per degree of freedom. The Hagedorn behavior makes the temperature stay constant at the Hagedorn temperature of the system $\beta=\log(2)$ for various values of the energy. In that sense, it is essentially a first order transition. 
Because of this, we have to study the system in the microcanonical ensemble.
The exit of the Hagedorn part of the phase diagram requires the temperature to start increasing again at a specific value of the rescaled energy per degree of freedom $\epsilon= E/N^2$.
We need to study when this happens.

 If we also take into account the charge, a phase transition would indicate that there will be a curve in $\epsilon, q$ where some thermodynamic quantities are discontinuous. That phase transition curve denotes the transition from the {\em partial deconfinement} phase to the {\em fully deconfined phase}.
Our goal in this section is to argue precisely how that phase transition appears in the counting of states done more carefully at large but finite $N$.

\subsection{State counting with Young tableaux}

Let us again start with the problem of counting states in the model we have described. 
The Hilbert space without constraints is described by the occupation numbers of the $2N^2$ harmonic oscillators. We can call these $(a_X^\dagger)^i_j$ and $(a_Y^\dagger)^i_j$. They have matrix indices, both an upper and a lower index. Generically, to build a state, one needs to contract the upper indices and lower indices. A naive counting of states is done in terms of traces that implement these contractions. However, the states created this way are not orthogonal states at finite $N$. At some point not only are the multi-traces not orthogonal, but one is also overcounting: there are relations. The traces are useful as algebraic generators of the states. Short traces are also simple observables that can be evaluated in a complicated state. In holography, these would represent excitations on top of a background.

Finding the orthogonal basis of states is not automatically easy. For a single matrix model, this is 
done using characters \cite{Corley:2001zk} and the representation is in terms of Young diagrams.
For more than one matrix, one can choose a basis of 
restricted Schur functions \cite{Balasubramanian:2004nb} (see also \cite{deMelloKoch:2007rqf,Brown:2007xh,Bhattacharyya:2008rb,Bhattacharyya:2008xy}), or  one can also find a double coset ansatz for writing explicit states \cite{deMelloKoch:2012ck}.

To understand this, the system is free. This means one can actually do rotations on the upper and lower indices of the $X$ and $Y$ independently of each other. 
We would then have a $U(N)^2$ symmetry of upper indices and lower indices separately for $X$ and another one such for $Y$. Basically, the starting symmetry is larger than $U(N)$, but only one $U(N)$ is gauged. It is convenient to use the extra symmetry to construct states,
By symmetry here, we mean that the $U$ act by unitary transformations on the Hilbert space. Therefore, states in different representations of the symmetry are orthogonal.
This is the idea behind the restricted Schur constructions. It is also a convenient way to analyze more general quiver theories ( see \cite{Pasukonis:2013ts,Berenstein:2015ooa}).
It is convenient to classify states in the full Hilbert space, including non-singlet states by the representation content under the $U(N)^4$ symmetry. 
The final $U(N)$ symmetry that we gauge sits in a diagonal of this $U(N)^4$. It acts on upper indices as a fundamental, and on lower indices as an antifundamental.
Thus, the $U(N)^4$ content also keeps track of the $U(N)$ gauge symmetry that we want to gauge in the end. 

The idea to represent the states is that the $X$ oscillators commute. To decompose into representations of $U(N)$ one symmetrizes or antisymmetrizes in the upper indices according to a Young diagram. We do the same with the lower indices. Notice that a permutation of two $X$ causes a permutation of both the upper and the lower indices that they carry. This is a commutative operation in the algebra of raising operators. A permutation of the upper indices can therefore be undone in the lower indices by this mechanism. This means that the Young diagram of the upper indices (the symmetry properties under permutations) is the same as the Young diagram of the lower indices \footnote{For fermions, the opposite is true \cite{Berenstein:2004hw}. See also \cite{Berenstein:2019esh} and references therein. Therefore the two Young diagrams are  transposed between upper and lower indices.}.  We can now do the same with the $Y$ oscillators. 

We now want to be more mindful of the four $U(N)$ symmetries. These will be called  $U(N)_{X,U}$,$ U(N)_{X,L}$,
$U(N)_{Y,U}$, $U_{Y,L}$, where we distinguish upper and lower indices by $U,L$.
We can organize the information we have collected so far by saying that 
we have four Young diagrams $\Upsilon_{X,U}= \Upsilon_{X,L}$ and $\Upsilon_{Y,U}= \Upsilon_{Y,L}$ and these are paired (identical between upper and lower indices of $X$ and $Y$ respectively).
Each of these is associated with an irreducible representation of $U(N)$. We now want to collect all the upper indices together.  Because the $U(N)$ we want to gauge acts the same on the upper indices of $X$ and $Y$, 
the main observation is that the upper indices transform as elements of a tensor product representation $R(\Upsilon_{X,U})\otimes R(\Upsilon_{Y,U}) $ with respect to this diagonal $U(N)$.
We decompose these into irreducible representations of the  diagonal action.
If we take two representations $R_1, R_2$, we have that 
\begin{equation}
R_1\otimes R_2 = \oplus_{R_3} c^{R_3}_{R_1,R_2} R_3
\end{equation}
where the $c^{R_3}_{R_1,R_2}$ are the multiplicities of the irreducible representation $R_3$ appearing in the product. These are known as Littlewood-Richardson coefficients. 
Now we do the same with the lower indices. This results in a different representation appearing on the lower indices, which we call $\tilde R_3$, with multiplicity $ c^{\tilde R_3}_{R_1,R_2}$.

The upper indices of $X$ transform as the fundamental with respect to the diagonal group $U(N)$ we are gauging and the lower indices transform in the conjugate representation. To make a singlet $R_3\otimes \bar {\tilde R}_3$ needs to contain a singlet. This can only occur if the Young diagrams of $R_3$ and $\tilde R_3$ are the same, and the multiplicity is one. 
Now, the upper indices have an additional degeneracy of $ c^{R_3}_{R_1,R_2}$ and the same is true for the lower indices. These need to be multiplied when we are counting states.
We find therefore that the partition function at fixed $n_x, n_y$ requires us to choose a young diagram for $X$ with $n_x$ boxes, a young diagram for $Y$ with $n_y$ boxes and a young diagram for the product representation, which must by necessity have $n_x+n_y$ boxes.
The total number of states is then a sum over all the representation choices obtained this way and counted with degeneracies
\begin{equation}
    N(n_x,n_y) = \sum_{ \nu=\Upsilon(n_x),\mu=\Upsilon(n_x), \sigma=\Upsilon(n_x+n_y)} 
    (c^\sigma_{\mu\nu})^2
\end{equation}
This result also appears in this form in \cite{Collins:2008gc,Mattioli:2016eyp} (see also \cite{deMelloKoch:2012sie,Ramgoolam:2016ciq}). There are other ways of generating the states using 

Two important observations are in order. First, if the Young diagram $\sigma=R_3$ has more 
than $N$ rows, then we do not count it, as it is not an allowed representation of $U(N)$. In that case, we set the corresponding $c^\sigma_{\mu\nu}$ to zero. 
Second, the Littlewood-Richardson coefficients are otherwise independent of $N$. 
This means that at finite $N$ and infinite $N$ the numbers $c^\sigma_{\mu\nu}$ are the same if they are allowed.  As a corollary, the counting of states at finite $N$ and infinite $N$ agree if the total number of boxes $n_x+n_y\leq N$. The partition function given by equation \ref{eq:part}, interpreted combinatorially in terms of these sums of squares of Littlewood Richardson coefficients is also known in the mathematics literature, a result that is attributed to Harris and Willenbring \cite{harris2014sums}.

\subsection{The typical Young tableaux}

We have two results concerning the counting of states. First, we have the infinite $N$ counting and we also have the finite $N$ counting, whose essential constraint is that all the Young tableaux $\Upsilon$
must be allowable for $U(N)$. If we combine both results, we get that when both countings are allowed then
\begin{equation}
    N(n_x,n_y) = \sum_{ \nu=\Upsilon(n_x),\mu=\Upsilon(n_x), \sigma=\Upsilon(n_x+n_y)} 
    (c^\sigma_{\mu\nu})^2\simeq \exp( \beta_q(n_x+n_y))
\end{equation}
At this stage, we want to ask what Young diagrams dominate the sum and how large do the 
$c^\sigma_{\mu\nu}$ become. Basically, we want to ask if maximizing over $c^\sigma_{\mu\nu}$ and effectively reducing the problem to one term is sufficiently representative of the entropy or not.
If the answer is yes (a statement that we will argue later),  we can then study how the shape of the dominant Young diagrams behaves as we take $n_x+n_y$ large. The main idea we want to advance is that if $\sigma$ is the dominant shape  and it is allowed for $U(N)$, then for all intents and purposes the entropy at finite $N$ and infinite $N$ at energy $E=n_x+n_y$ are the same. Their difference in entropy will be small and suppressed. If the shape is not allowed for $U(N)$, then the state counting for $U(N)$ and $U(\infty)$ is substantially different at energy $E$. The energy at which the dominant shape for $E=n_x+n_y$ ceases to be allowed is then associated with a change of thermodynamic behavior away from the result at infinite $N$. This is the critical point in $E$ that we are looking for.

\subsubsection*{Large Littlewood Richardson coefficients}

So far, we have used group theory to argue that the counting of states can be done by summing over triples of Young diagrams, with $n_x$, $n_y$ and $n_x+n_y$ boxes. How many of these triples are there? 
The number of young diagrams with $n_x$ boxes is given by the partitions of $n_x$. The same is true for $n_y, n_x+n_y$. 
The asymptotic number of partitions at large $n_x, n_y$ (without any constraints) is
\begin{equation}
    P(n_x, n_y, n_x+n_y) \sim \exp\left(\pi \sqrt{2n_x/3}+\pi \sqrt{2n_y/3}+\pi \sqrt{2(n_x+n_y)/3} \right) .
\end{equation}
This means that the maximum possible entropy associated with this sum (if all terms are the same)
scales like
\begin{equation}
    S_{\# terms} \sim \pi \sqrt{2n_x/3}+\pi \sqrt{2n_y/3}+\pi \sqrt{2(n_x+n_y)/3} \ll \beta_q (n_x+n_y)
\end{equation}
which is much smaller than the entropy of the system. After all, they scale like $\sqrt n$, rather than $n$. 
In essence, we find that the entropy is not concentrated on the number of partitions. Instead, we can find the following inequality
\begin{equation}
    \log(P(n_x, n_y, n_x+n_y) {\max_{\mu,\nu,\sigma} (c^\sigma_{\mu,\nu}
    })^2 )> S \sim \beta_q (n_x+n_y)
\end{equation}
Also, if we reduce the sum to the one term that maximizes the Littlewood Richardson coefficient, we find that
\begin{equation}
   \log({\max_{\mu,\nu,\sigma} (c^\sigma_{\mu,\nu}
    })^2 )< S
\end{equation}

Combining these two, we find that 
\begin{equation}
\log(\max_{\mu,\nu,\sigma} (c^\sigma_{\mu,\nu}
    )^2)\sim \beta_q(n_x+n_y) - O(\sqrt n)
\end{equation}
We conclude that the term with the maximum Littlewood Richardson coefficient has an entropy associated with it that is roughly equal to the thermodynamic entropy of the system, up to subleading corrections that can be treated as a small perturbation. The basic claim we make is that the term with the maximum Littlewood-Richardson coefficient is sufficiently representative.

Our next problem is to find at large $n_x, n_y$, what is the shape of the Young diagram that maximizes the Littlewood-Richardson coefficient, if there is such a shape. This is a well-known problem in combinatorics.  
We will here quote the main result of \cite{PAK201944} on the asymptotic behavior of the shape associated with the maximum Littlewood Richardson coefficient. 
The shape of the asymptotic young diagram is known as the VKLS shape. 

To understand what this shape does, let us recall the dimension of the
representation associated with a Young diagram. This is given by taking a product of labels associated with each box and dividing by the hook lengths. The labels of the boxes are as follows, shifted by $N$.
\begin{equation}
\ytableausetup{mathmode, boxframe=normal, boxsize=2em}
\begin{ytableau}
 0 & 1 & 2 & 3& \dots \\
 -1 & 0& 1& \dots\\
 -2& -1 & 0& \dots\\
 \vdots& \ddots& \ddots\\
\end{ytableau}
\end{equation}
They start by $0$ in the $(1,1)$ corner and add one when moving to the right and substracting one going vertically down. Basically, it is $i-j$, where $i$ is the horizontal label, and $j$ is the vertical label counting from he top.
Let us call the label of the $(i,j)$ box $L_{i,j}$
The dimension of the representation is given by
\begin{equation}
d_{\nu}= \prod_{i,j} \frac{(N+ L_{i,j})}{h_{i,j}}
\end{equation}
where $h_{i,j}$ is the hook length of the $(i,j)$ box.
When we consider the large $N$ limit, we have that 
\begin{equation}
   \tilde d_\nu= \lim_{N\to \infty} \frac{d_{\nu}}{N^{|nu|}} = \prod_{i,j} \frac 1{h_{i,j}}
\end{equation}
Roughly stated, the normalized size of the representation is the inverse product of the hooks of the Young diagram.  The VKLS shape is the asymptotic
shape that maximizes the normalized dimension $\tilde d_\nu$ when we take large values, $|\nu|\to \infty$. 
Taking logarithms, we find that 
\begin{equation}
\log(\tilde d_\nu)= -\sum_{i,j} \log(h_{i,j})
\end{equation}
To maximize $d_\nu$, we must minimize the sum $F=\sum_{i,j} \log(h_{i,j})\propto \int dx dy \log(h_{x,y})$, which can be represented as an integral. Since the number of boxes is fixed, we can choose the area of the $(x,y)$ plane to be fixed and be equal to one. The VKLS shape is the shape of the region that minimizes the functional $F$ at fixed area, equal to one.
The shape is described as follows. Consider  the region in between the two curves 
\begin{eqnarray}
    f(s)&=&\frac{2 \left(\sqrt{2-s^2}+s \sin ^{-1}\left(\frac{s}{\sqrt{2}}\right)\right)}{\pi } \\
    \tilde f(s)&=&|s|
\end{eqnarray}
in the interval $s\in( -\sqrt 2,\sqrt 2)$. If we think of the curve given by $|s|$ as the labels of the rows and columns of the Young diagram, the curve $f(s)$ rotated so that it lies in the lower right quadrant is the VKLS shape.
Importantly, the $f(s)$ curve intersects the $|s|$ curve at $s=\pm\sqrt{2}$. The distance from the origin in geometric units is $2$.
In the asymptotic calculation of \cite{PAK201944}, all three shapes have the VKLS shape, properly scaled to the corresponding number of boxes.

\begin{figure}[ht]
\includegraphics[width= 10 cm]{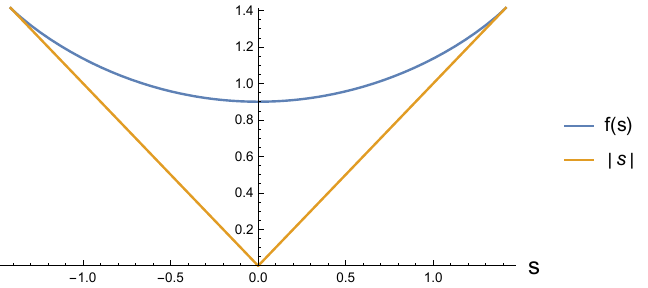}
\caption{A schematic representation of the VKLS shape. Boxes of the Young diagram must fill the corner defined by the function $|s|$ with sides parallel to the $|s|$ lines.}\label{fig:VKLS.pdf}
\end{figure}

The VKLS shape, as described above, is depicted in figure \ref{fig:VKLS.pdf}.

We need to convert the area to the correct number of boxes to restore units:  the area is $n_x+n_y$ rather than one. 
The length of the legs must be scaled by $\sqrt{n_x+n_y}$ to accomplish this.
Therefore, the depth of the VKLS shape Young diagram in proper units is $2\sqrt{n_x+n_y} \leq N$, and it must be bounded by $N$ as that specifies the maximum allowed depth of the Young diagram columns. We find that the VKLS shape is allowed only if $E=n_x+n_y\leq N^2/4$.

Our prediction for the transition from partial deconfinement to full deconfinement based on this argument is that it occurs exactly at energy $E=N^2/4$, regardless of the value of $q$. 
The value $N^2/4$ is also reported in \cite{O'connor}, by using different means.
This is the asymptotic large $N$ statement, so there can be corrections that are subleading in $N$ that we can not account for from the arguments above.
To test this statement, we do numerical calculations to see if the change of behavior occurs at fixed energy per degree of freedom $\epsilon=E/N^2= 1/4$.

From our perspective, the partially deconfined gauge group has size $2\sqrt{n_x+n_y}$ and the confined portion is $SU(N-2\sqrt{n_x+n_y})$. This is done by looking at the number of rows less than $N$ that are empty in the Young diagram. The definition is as in \cite{Berenstein:2018lrm}. In this paper, we can actually quantify this property at large $N$. In this setup, we do not have access to the characterization of states in terms of the distribution of eigenvalues of the  Polyakov loop, as in \cite{Hanada:2018zxn}, or in terms of the absolute value of the Polyakov loop.

\section{Numerics and the phase transition}\label{sec:numeric}

In this section, we are providing numerical evidence that the reasoning above is correct. The process is twofold. First, we wish to calculate the maximum Littlewood Richardson coefficients and compare them to the hook length formula. We wish to check that these coefficients are maximized sharply for the lower values of the hook length. Secondly, we need to compare different $N$ in a meaningful way. The simplest way to do so is to notice that large $N$ scaling requires that both $E,S\sim N^2$, so that we need to base our calculations on rescaled energy $\epsilon=E/N^2$ and rescaled entropy $s= S/N^2$. Since $S\propto E$ in the Hagedorn region, it is convenient to use the rescaled free energy $F/N^2= E/N^2- T S/N^2 = \epsilon - T s$, which  vanishes at large $N$ for $\epsilon<\epsilon^*$, the energy of the phase transition. At least in principle, this provides a convenient parameter to distinguish the two phases $F/N^2=0$ and $F/N^2\neq 0$. This parameter changes continuously at the phase transition.

When doing calculations at finite $N$, there should be finite $N$ corrections on top of these that we can not determine directly from the limit shape without extra input. Roughly stated, the VKLS curve is an approximation to the rugged edges of the Young diagrams. Because the 
curve becomes tangent to $|s|$ at the edge of the distribution, how to treat the edge can affect the size of the Young diagram versus the edge of the VKLS shape. This can be an effect that is much larger than order $1$ but necessarily much less than $\sqrt{n_1+n+2}$, the naive size of the shape. At this stage, this is a systematic error that affects how quickly the systems converge to the large $N$ result at very moderate $N\simeq 4$--$ 7$, where we  will be doing our calculations. To estimate roughly, the Young diagram can cover completely the VKLS curve, or instead be  completely covered by the VKLS curve. The difference in area between the two is of order the length of the edge of the diagram. This scales like $\sqrt{n_1+n_2}$. Since $n_1+n_2 \simeq O(N^2)$, this difference is of order $N$. We therefore expect that the transition occurs at $n_1+n_2 = N^2/4+ O(N)$. The additional piece must be positive, as for $n=4,5,6$ the number $N^2/4$ is still small, especially if compared to the maximal depth of the Young diagram, which has roughly the same size.

For the first part, we check numerically that the problem that gives rise to the VKLS shape is
sound in the regime of parameters we are analyzing. We compute for fixed $E=n_x+n_y$   the distribution of the  Littlewood-Richardson coefficients as a function of the hook length formula of the large Young diagram,  at $E=n_x+n_y$ boxes with both species equal to each other  $n_x=n_y$.
This is done by using the {\text lrcalc} package in Sage. We generate all Young diagrams for $n_x$ boxes and $n_x+n_y$ boxes. We make sure that 
 the maximum depth of the Young Diagram is fixed at $N=5,6,7$ to compare different values of 
$N$. We compute the distributions by iterating over these choices.

\begin{figure}[ht]
    \centering
    \includegraphics[width=6 cm]{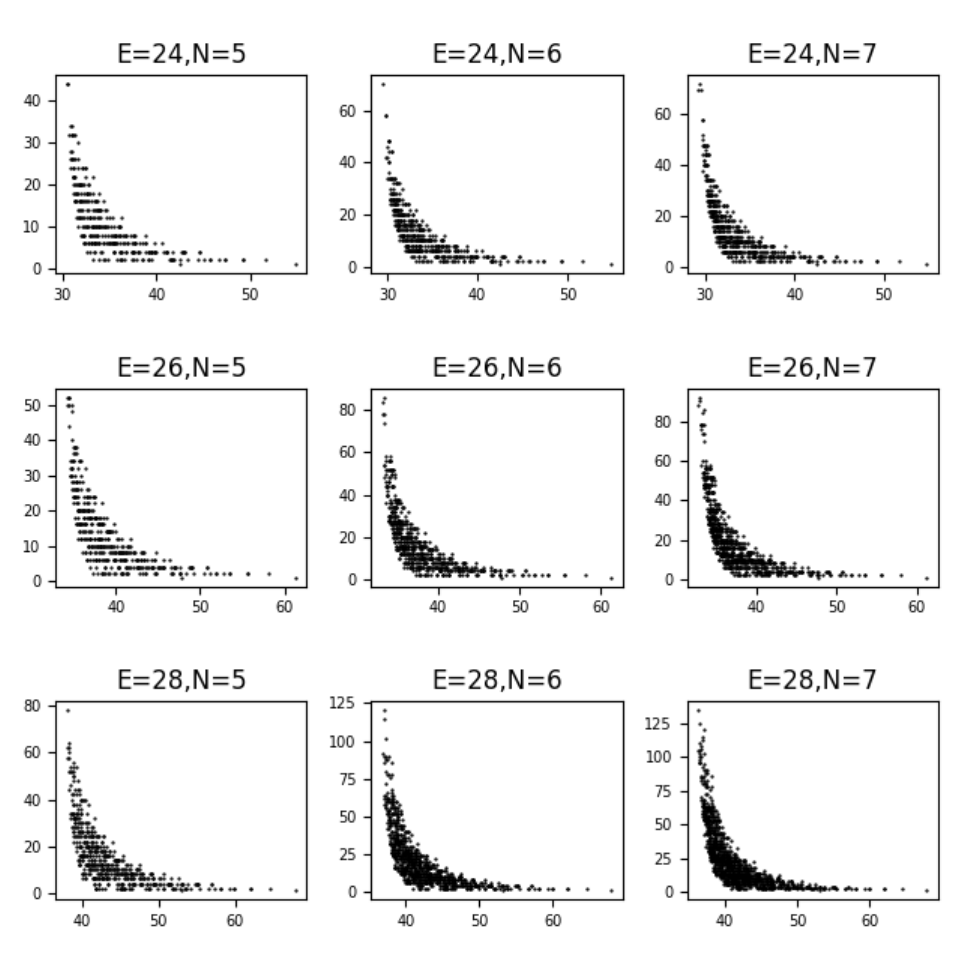}
    \includegraphics[width=6 cm]{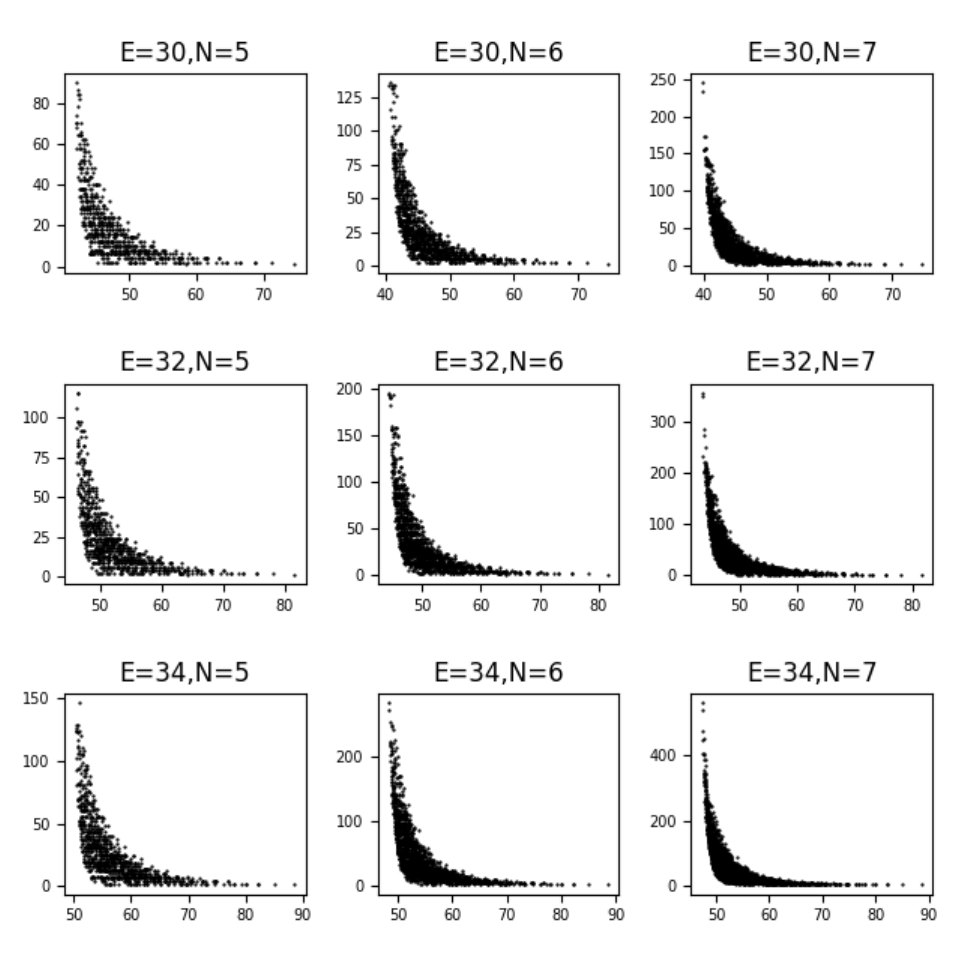}
     \caption{The x-axis is the logarithm of the hook product of $\nu_k$, and the y-axis is the Littlewood-Richardson coefficient. This plot includes systems with total energy $E=24$ to $E=32$, and the Young diagrams represent $SU(N)$ with $N=5,6,7$}
    \label{fig:LR_vs_hook(24-28).pdf}
\end{figure}

This is depicted in figure \ref{fig:LR_vs_hook(24-28).pdf}. We clearly see that the maximum Littlewood-Richardson coefficient is peaked at low values of the hook length formula. 

We also point out that as we increase the energy, the value of $N$ at which the coefficient  distribution peaks and saturates grows and the hook length formula moves towards the left (decreases). 

\subsection{Free energy}
\begin{figure}[ht]
    \centering
    \includegraphics[width=0.7\textwidth]{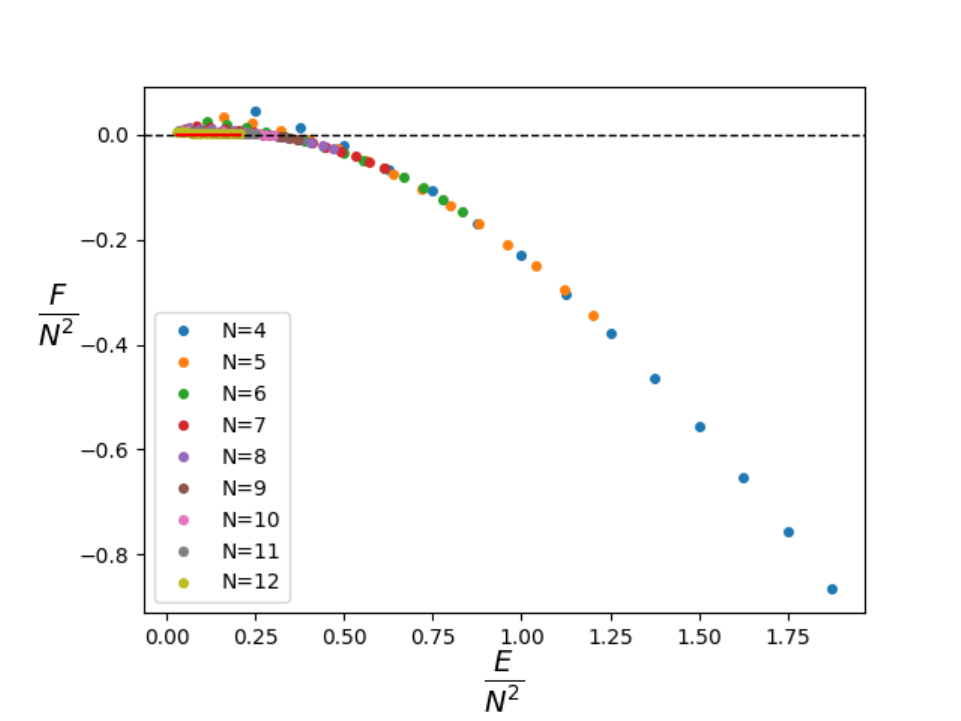}
    \caption{The rescaled free energy $\tilde{F}$ is plotted against the scaled energy $\tilde{E}$ for various $N$ ranging from $4$ to $12$. The zero point is indicated by the red line.}
    \label{fig:Free_energy_scaled_q_0.pdf}
\end{figure}

The next step is to compute the free energy. We have argued 
that in the absence of charge, before the transition the scaling of the entropy is given by:
\begin{equation}
    S = E\ln2
\end{equation}
The effective temperature is then:
\begin{equation}
    \beta_{eff} = \frac{\partial S}{\partial E} = \ln 2
\end{equation}
It is easy to see that $F=0$ at this temperature. After the transition, however, both temperature and entropy scale as the power laws in energy, and similarly for the free energy. we compute the free energy summing over all allowed states, not just the one that maximizes the Littlewood Richardson coefficient. 
The temperature is computed in the microcanonical ensemble by finite differences $T\equiv (\Delta S/\Delta E)^{-1}$, at fixed $q=0$. This results in some dispersion relative to the $\beta_{eff}=\log(2) $ from large $N$ when we do it at finite $N$. We normalize both the energies and the free energy by dividing by $N^2$, to check convergence for large $N$. Since the Littlewood 
Richardson coefficients are hard to compute, in practice we are restricted in energy $E\leq 34$. For $N=12$ (the maximum depth we can compute at), we have $N^2/4=36$, which is larger than the maximum energy where we did our computations. Therefore the data at this level is below the expected transition point.
The figure
Fig \ref{fig:Free_energy_scaled_q_0.pdf} shows that the rescaled free energy $\tilde{F}$ versus the rescaled energy $\tilde{E}$ collapses at large $E/N^2$ and that deviations start to appear close to  $E^*\simeq 0.25 N^2$. At larger $N$, the curve flattens to zero  below $E^* \simeq 1/4 N^2$.
One can see that $\tilde{F}$ remains relatively flat and close to zero all the way up to $\tilde{E}\approx 0.25\approx \tilde{E_*}$, as we conjectured earlier. 

\begin{figure}[ht]
    \centering
    \includegraphics[width=0.7\textwidth]{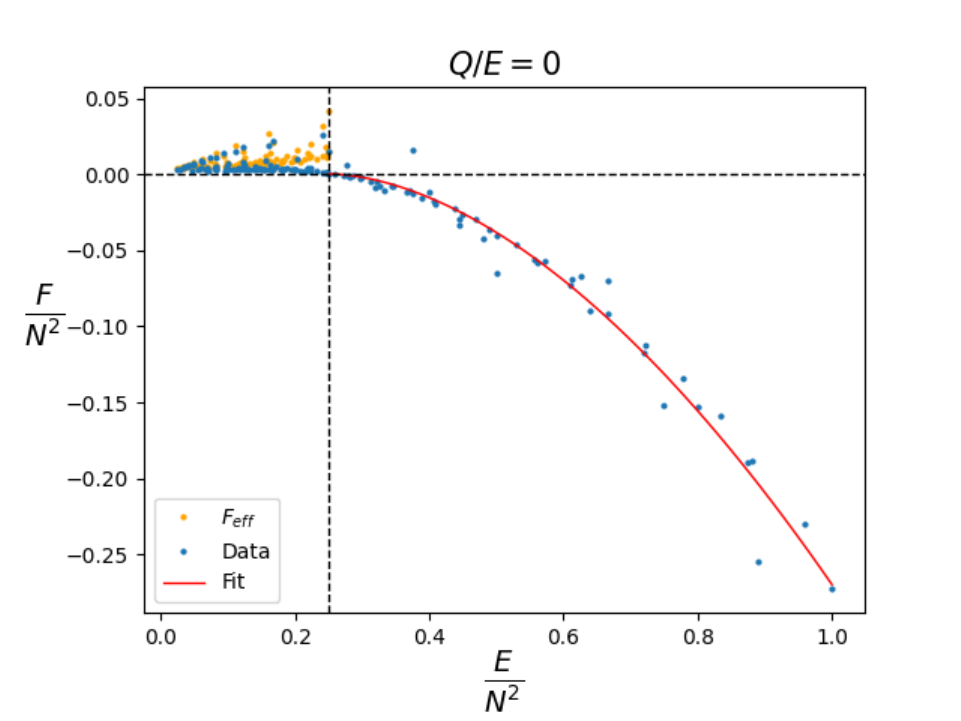}
    \caption{The plot shows the rescaled free energy $\frac{F}{N^2}$ agains the rescaled energy $\frac{E}{N^2}$. The simulated data(blue) are computed by taking the temperature as the discrete derivative $\partial S/\partial E$. $F_{eff}$ is computed by taking the temperature to be the inverse of $\beta_{eff}$. The horizontal dashed line corresponds to $F/N^2=0$, and the vertical dashed line corresponds to $E/N^2=0.25$. A power law fit is performed for $\tilde{F}$ corresponding to $0.25<\tilde{E}<0.5, \ N = 3,4,5,...,13$. }
    \label{fig:free_energy_fit_atratio_0.pdf}
\end{figure}

We also check to see if we have non-trivial critical exponents at the transition, assuming that $E^*/N^2= \epsilon^*=1/4$ in figure \ref{fig:free_energy_fit_atratio_0.pdf}. 

In the figure we include two determinations of the free energy for $\epsilon^*<1/4$. We compute the free energy with the temperature determined by finite differences, and compare it to the free energy assuming that $\beta=\log(2)$ is fixed. The energy relative to the conjectured transition point is $\tilde \epsilon= \epsilon -1/4 = \epsilon-\epsilon^*$.
The best value of the fit is 
\begin{equation}
    \tilde{F} \propto -\tilde{\epsilon}^{1.77}\sim- (\epsilon-\epsilon^*)^{2-1/4}
\end{equation}
which seems to show a non-trivial critical exponent. Given our data, we are choosing a simple rational number as a stand-in for the exponent. 
The seemingly strange straight lines in the figure are actually the same data point assuming different values of $N$: both the free energy and the energy are rescaled by the same amount $1/N^2$. This becomes more obvious when the coloring of figure 
\ref{fig:Free_energy_scaled_q_0.pdf} is used to parse the data. The second plot zooms closer into the region $\epsilon \simeq 0.25$. We also see that $F/N^2\simeq 0$ is well supported below $\epsilon=1/4$.

Using the relations $d\epsilon = T d s $ and $f=F/N^2= \epsilon - T s$, we find that 
$df \simeq -dT s = -dT(s-s^*)- s^* dT  $. Since $s=s^*$ at the transition, the term with $(s-s^*)$ is suppressed. Instead, we find that $dT/d\epsilon\propto (\epsilon-\epsilon^*)^{2.77}$, so that we should have $T= T^* + \alpha (\epsilon-\epsilon^*)^{1.77} $ near the transition. The relation between $T-T^*$ and $\epsilon-\epsilon^*$ shows a non-trivial critical exponent, where $\epsilon-\epsilon^* \simeq (T-T^*)^{0.56}$. Since the power law is less than one, the specific heat itself diverges with a non-trivial exponent, signaling a weak first order transition. This is very similar to the critical exponents found in \cite{Pisarski:2012bj}

It should be interesting to derive these exponents directly from the change of shape of the Young diagram..

\subsection{Charge dependence}

\begin{figure}[h]
    \includegraphics[width=7 cm]{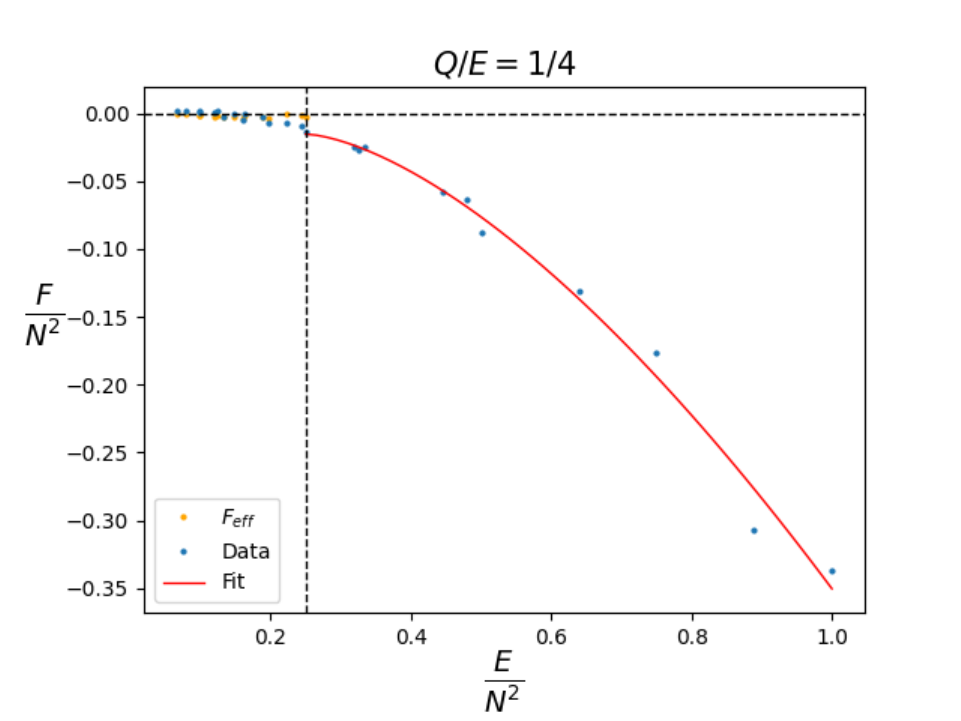}\includegraphics[width=7 cm]{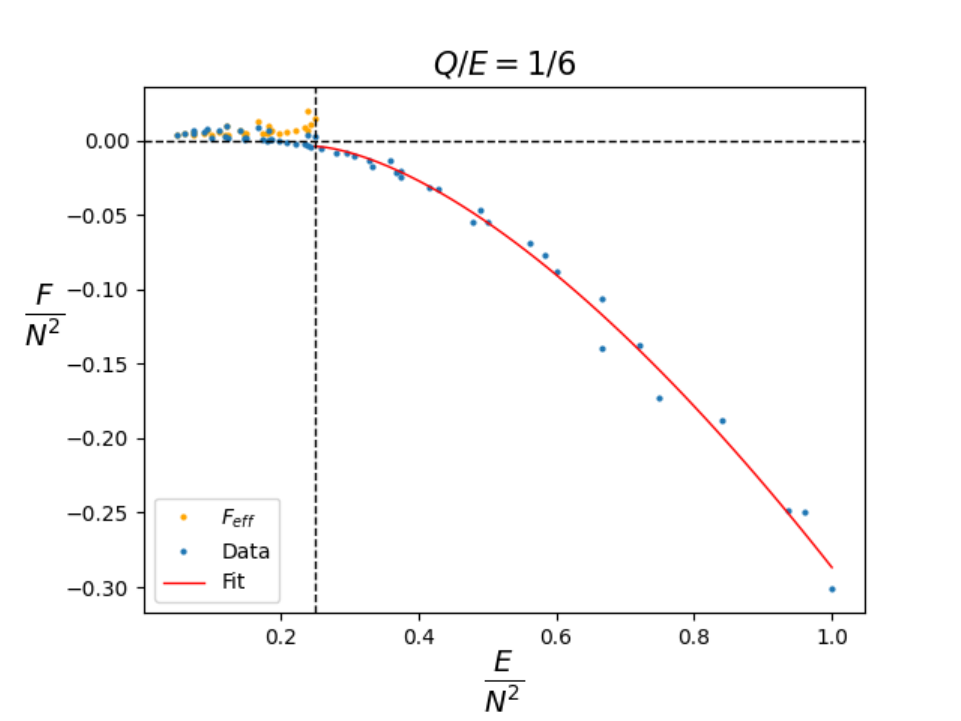} \includegraphics[width=7 cm]{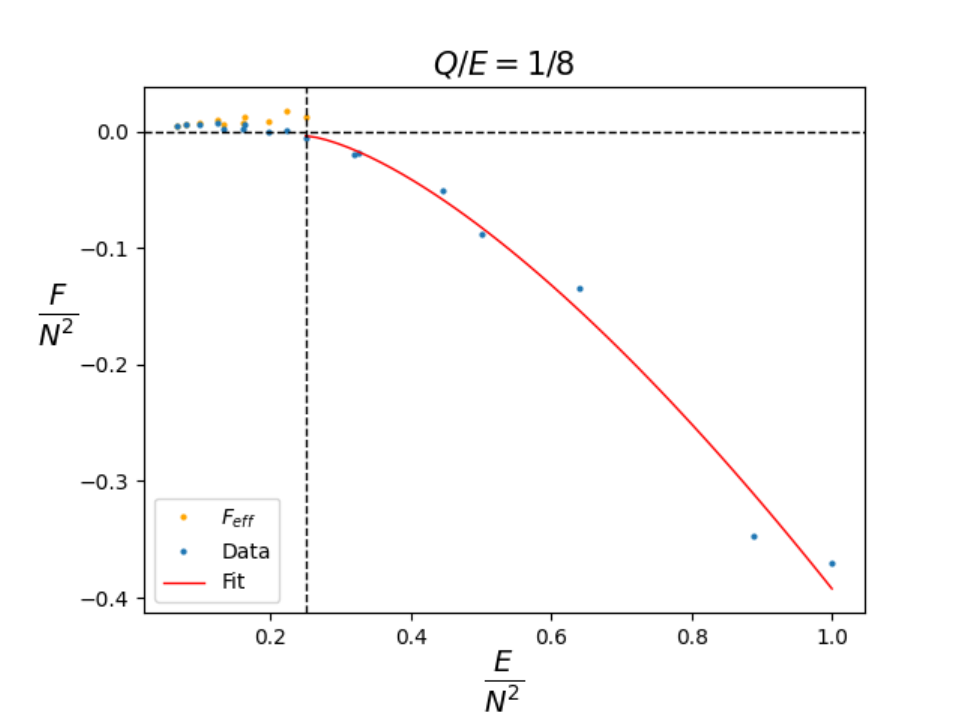}
    \caption{The plot of $\tilde{F}$ versus $\tilde{E}$ for various values of $\frac{Q}{E}$, following the same conventions as figure \ref{fig:free_energy_fit_atratio_0.pdf}. Fits to power laws are made. }
    \label{fig:1_over_4_e_lessthan_1.pdf}
\end{figure}

Our arguments in general require that the phase transition always occur at energy $E=N^2/4 $.
We have found evidence in the case $Q=0$ that this is the case. We now want to do that at $Q\neq 0$. To get a proper limit, we keep $q$ fixed (equivalently $Q/E$ fixed). This ratio is
$q=(n_1-n_2)/( 2(n_1+n_2))$. If we want for example $n_1=2 n_2$, this corresponds to $q= Q/E= 1/6$, and $n_1= 3 n_2$ corresponds to $q= 1/4$, whereas $3 n_1= 5n_2$ is $q=1/8$.
These must be done at energies that are multiples of $3,4,8$ respectively. The number of data points we can actually compute is more sparse with multiples of $4,8$, so it is less reliable,

The same information as in \ref{fig:free_energy_fit_atratio_0.pdf} can be plotted at different $Q/E$. We get the results in figure \ref{fig:1_over_4_e_lessthan_1.pdf}.
The figures for different $Q/E$ all support the idea that the phase transition occurs exactly at $E/N^2= \frac 14$ and the plots are qualitatively very similar. For $Q/E= 1/4$ the charge is getting large and closer to the maximum value $Q/E=1/2$. 

Naive power fits with a shift do not show a universal behavior, other that the critical exponent being larger than $1$. The data is also sparse. The best data point is at $q=1/6$ and at a relatively low $N$. This is shown in \ref{tab:exp}. 
\begin{table}[h!tb]
\centering
\begin{tabular}{|c|c|c|c|c|}
\hline
    $q=Q/E$     & 1/4 & 1/6 & 1/8&0 \\ \hline
Power    &   1.95  &  1.56   & 1.46 &1.77   \\ \hline
$F(0)$   &  -0.026   &   -0.004  &    -0.004&0 \\ \hline
\end{tabular}
\caption{Power law fits to the exponent}\label{tab:exp}
\end{table}
The $\chi^2$ is best for $q=1/6$, but given the variance of all the answers, we need more data to make a more definite statement. The question of if the fit is good or not at this stage has too many systematic errors to put a proper error bar on it. The main reason is that we do not know if the range is small enough for the power law fit to be dominated by the first non-trivial term.
Asymptotically, the temperature becomes linear in $\epsilon$  and the free energy must scale like $-\epsilon\log(\epsilon)$. The cutoff $\epsilon\simeq 1$  might be too large a cutoff for larger $q$. We also don't know if $N$ is large enough for finite $N$ corrections
to be unimportant or not. This requires much more data at high energy \footnote{The Littlewood Richardson coefficients are asymptotically hard to compute \cite{narayanan2005computation}, so in our computations, we have had to make a table of all possibilities. The limits we have here correspond to what could be reasonably computed with a laptop.}.

\section{Conclusion}\label{sec:con}

In this paper, we have presented both theoretical and numerical evidence that the transition from partial deconfinement to full deconfinement can be understood simply in terms of counting  of states for  the free gauge matrix model based on Young diagrams. These have a typical shape, and when the typical shape, scaled to the number of boxes reaches the maximum allowed depth of the Young diagrams, the transition takes place. Before the transition,  the shape is independent of the charges. We presented numerical evidence that this occurs exactly at the place where this counting suggests.  
At the exit point,  the large $N$ free energies stops being zero.  There are non-trivial critical exponents on the exit side of the Hagedorn region of the microcanonical phase diagram, which verifies with our methods that the transition is weakly first order. 

The claim we are making is that the transition from partial deconfinement to deconfinement corresponds to a change in the typical shape of the Young diagram. To the extent that the shape of the Young diagram can be also considered as a geometric object, the transition as we describe above is stating that there is a geometric interpretation of the transition (a geometric order parameter), which is different from the description of the transition in terms of the absolute value of the Polyakov loop that has been used in other works. How to relate our observations with the VKLS shape to the Polyakov loop is beyond the scope of the present paper, but it should be an interesting avenue of exploration. Both of these approaches are very different in how one deals with the physical questions.

The problem of the shape of the Young diagram seems to be intimately related to counting states. If one replaces the problem of counting states with Young diagrams with the problem of counting states with traces, the transition occurs when the number of relations between traces  competes with the number of states to the point that there are large cancelations and the entropy decreases substantially from what infinite $N$ would dictate. Basically, traces are becoming very redundant.
If we equate entropy with information, we can say that this is a transition on the information content of the state. This is also suggestive of a closer connection with black holes as the entropy can be computed geometrically for black holes. 
Notice that this description is an alternative point of view to the change in the expectation value of the Polyakov loop variables, which relates the problem to a change in the  distribution of eigenvalues of the gauge field.  

It is clear that our techniques work also in cases of more matrices or in systems with fermions instead of bosons.
One then needs to consider more young diagrams or different combinations of them, but with our methods, the computations again require maximizing products of Littlewood Richardson coefficients.
All of these will give rise to variations of the combinatorial problem that leads to the VKLS shape, as described in this paper. It is this effective shape that is controlling the transition in  all these setups.

Also, given the information about the phase transition that can be learned from physics calculations, one should keep in mind that the physics intuition may also bear some fruit in the study of  the estimation of Littlewood Richardson coefficients beyond the VKLS regime. This is an important combinatorial problem in its own right,

It is obviously interesting to ask how to translate combinatorial information about Young diagrams into computations of other observables in the matrix model. 
As a case in point, for  the one matrix model and because of its relations to half BPS states in ${\cal N}=4$ SYM, a collection of such methods have been understood in \cite{Berenstein:2017abm,Berenstein:2017rrx} (see also \cite{Balasubramanian:2018yjq}). It would be interesting to understand similar statements in this setup. At least in principle, since we know how to write the $SU(N)$ generators for $X,Y$ separately, information on the shape of the Young tableaux can be obtained by building the Casimir operators of the different $SU(N)$ groups that are not gauged. Hopefully, this will lead to an improvement in the understanding of correlators for these states and how these are modified  when changes occur in the typical Young diagram. That should lead to an interesting determination of the critical behavior near the partial deconfinement to deconfinement transition.

Ideally, because the VKLS states dominate the entropy in this case, the VKLS shape states could also dominate in cases where the theory is interacting with a non-trivial potential. In these cases, a microcanonical computation would be out of reach by direct methods. 
These interacting models are closer to black holes in that one would expect to have chaotic dynamics and satisfy the eigenstate thermalization hypothesis. Maybe they could even have negative specific heat. We are currently looking into these ideas.

\acknowledgments
D.B. would like to thank D. O'Connor, S. Ramgoolam for discussions and correspondence. D.B. research was supported in part by the International Centre for Theoretical Sciences (ICTS) while participating in the program - ICTS Nonperturbative and Numerical Approaches to Quantum Gravity, String Theory and Holography (code: ICTS/numstrings-2022/9). Research supported in part by the Department of Energy under grant DE-SC 0011702.
\bibliography{refs.bib}

\begin{thebibliography}{46}%
\makeatletter
\providecommand \@ifxundefined [1]{%
 \@ifx{#1\undefined}
}%
\providecommand \@ifnum [1]{%
 \ifnum #1\expandafter \@firstoftwo
 \else \expandafter \@secondoftwo
 \fi
}%
\providecommand \@ifx [1]{%
 \ifx #1\expandafter \@firstoftwo
 \else \expandafter \@secondoftwo
 \fi
}%
\providecommand \natexlab [1]{#1}%
\providecommand \enquote  [1]{``#1''}%
\providecommand \bibnamefont  [1]{#1}%
\providecommand \bibfnamefont [1]{#1}%
\providecommand \citenamefont [1]{#1}%
\providecommand \href@noop [0]{\@secondoftwo}%
\providecommand \href [0]{\begingroup \@sanitize@url \@href}%
\providecommand \@href[1]{\@@startlink{#1}\@@href}%
\providecommand \@@href[1]{\endgroup#1\@@endlink}%
\providecommand \@sanitize@url [0]{\catcode `\\12\catcode `\$12\catcode
  `\&12\catcode `\#12\catcode `\^12\catcode `\_12\catcode `\%12\relax}%
\providecommand \@@startlink[1]{}%
\providecommand \@@endlink[0]{}%
\providecommand \url  [0]{\begingroup\@sanitize@url \@url }%
\providecommand \@url [1]{\endgroup\@href {#1}{\urlprefix }}%
\providecommand \urlprefix  [0]{URL }%
\providecommand \Eprint [0]{\href }%
\providecommand \doibase [0]{http://dx.doi.org/}%
\providecommand \selectlanguage [0]{\@gobble}%
\providecommand \bibinfo  [0]{\@secondoftwo}%
\providecommand \bibfield  [0]{\@secondoftwo}%
\providecommand \translation [1]{[#1]}%
\providecommand \BibitemOpen [0]{}%
\providecommand \bibitemStop [0]{}%
\providecommand \bibitemNoStop [0]{.\EOS\space}%
\providecommand \EOS [0]{\spacefactor3000\relax}%
\providecommand \BibitemShut  [1]{\csname bibitem#1\endcsname}%
\let\auto@bib@innerbib\@empty
\bibitem [{\citenamefont {Witten}(1998)}]{Witten:1998zw}%
  \BibitemOpen
  \bibfield  {author} {\bibinfo {author} {\bibfnamefont {Edward}\ \bibnamefont
  {Witten}},\ }\bibfield  {title} {\enquote {\bibinfo {title} {{Anti-de Sitter
  space, thermal phase transition, and confinement in gauge theories}},}\
  }\href {\doibase 10.4310/ATMP.1998.v2.n3.a3} {\bibfield  {journal} {\bibinfo
  {journal} {Adv. Theor. Math. Phys.}\ }\textbf {\bibinfo {volume} {2}},\
  \bibinfo {pages} {505--532} (\bibinfo {year} {1998})},\ \Eprint
  {http://arxiv.org/abs/hep-th/9803131} {arXiv:hep-th/9803131} \BibitemShut
  {NoStop}%
\bibitem [{\citenamefont {Hawking}\ and\ \citenamefont
  {Page}(1983)}]{Hawking:1982dh}%
  \BibitemOpen
  \bibfield  {author} {\bibinfo {author} {\bibfnamefont {S.~W.}\ \bibnamefont
  {Hawking}}\ and\ \bibinfo {author} {\bibfnamefont {Don~N.}\ \bibnamefont
  {Page}},\ }\bibfield  {title} {\enquote {\bibinfo {title} {{Thermodynamics of
  Black Holes in anti-De Sitter Space}},}\ }\href {\doibase 10.1007/BF01208266}
  {\bibfield  {journal} {\bibinfo  {journal} {Commun. Math. Phys.}\ }\textbf
  {\bibinfo {volume} {87}},\ \bibinfo {pages} {577} (\bibinfo {year}
  {1983})}\BibitemShut {NoStop}%
\bibitem [{\citenamefont {Sundborg}(2000)}]{Sundborg:1999ue}%
  \BibitemOpen
  \bibfield  {author} {\bibinfo {author} {\bibfnamefont {Bo}~\bibnamefont
  {Sundborg}},\ }\bibfield  {title} {\enquote {\bibinfo {title} {{The Hagedorn
  transition, deconfinement and N=4 SYM theory}},}\ }\href {\doibase
  10.1016/S0550-3213(00)00044-4} {\bibfield  {journal} {\bibinfo  {journal}
  {Nucl. Phys. B}\ }\textbf {\bibinfo {volume} {573}},\ \bibinfo {pages}
  {349--363} (\bibinfo {year} {2000})},\ \Eprint
  {http://arxiv.org/abs/hep-th/9908001} {arXiv:hep-th/9908001} \BibitemShut
  {NoStop}%
\bibitem [{\citenamefont {Aharony}\ \emph {et~al.}(2004)\citenamefont
  {Aharony}, \citenamefont {Marsano}, \citenamefont {Minwalla}, \citenamefont
  {Papadodimas},\ and\ \citenamefont {Van~Raamsdonk}}]{Aharony:2003sx}%
  \BibitemOpen
  \bibfield  {author} {\bibinfo {author} {\bibfnamefont {Ofer}\ \bibnamefont
  {Aharony}}, \bibinfo {author} {\bibfnamefont {Joseph}\ \bibnamefont
  {Marsano}}, \bibinfo {author} {\bibfnamefont {Shiraz}\ \bibnamefont
  {Minwalla}}, \bibinfo {author} {\bibfnamefont {Kyriakos}\ \bibnamefont
  {Papadodimas}}, \ and\ \bibinfo {author} {\bibfnamefont {Mark}\ \bibnamefont
  {Van~Raamsdonk}},\ }\bibfield  {title} {\enquote {\bibinfo {title} {{The
  Hagedorn - deconfinement phase transition in weakly coupled large N gauge
  theories}},}\ }\href {\doibase 10.4310/ATMP.2004.v8.n4.a1} {\bibfield
  {journal} {\bibinfo  {journal} {Adv. Theor. Math. Phys.}\ }\textbf {\bibinfo
  {volume} {8}},\ \bibinfo {pages} {603--696} (\bibinfo {year} {2004})},\
  \Eprint {http://arxiv.org/abs/hep-th/0310285} {arXiv:hep-th/0310285}
  \BibitemShut {NoStop}%
\bibitem [{\citenamefont {Harmark}\ and\ \citenamefont
  {Wilhelm}(2018{\natexlab{a}})}]{Harmark:2017yrv}%
  \BibitemOpen
  \bibfield  {author} {\bibinfo {author} {\bibfnamefont {Troels}\ \bibnamefont
  {Harmark}}\ and\ \bibinfo {author} {\bibfnamefont {Matthias}\ \bibnamefont
  {Wilhelm}},\ }\bibfield  {title} {\enquote {\bibinfo {title} {{Hagedorn
  Temperature of AdS$_5$/CFT$_4$ via Integrability}},}\ }\href {\doibase
  10.1103/PhysRevLett.120.071605} {\bibfield  {journal} {\bibinfo  {journal}
  {Phys. Rev. Lett.}\ }\textbf {\bibinfo {volume} {120}},\ \bibinfo {pages}
  {071605} (\bibinfo {year} {2018}{\natexlab{a}})},\ \Eprint
  {http://arxiv.org/abs/1706.03074} {arXiv:1706.03074 [hep-th]} \BibitemShut
  {NoStop}%
\bibitem [{\citenamefont {Harmark}\ and\ \citenamefont
  {Wilhelm}(2018{\natexlab{b}})}]{Harmark:2018red}%
  \BibitemOpen
  \bibfield  {author} {\bibinfo {author} {\bibfnamefont {Troels}\ \bibnamefont
  {Harmark}}\ and\ \bibinfo {author} {\bibfnamefont {Matthias}\ \bibnamefont
  {Wilhelm}},\ }\bibfield  {title} {\enquote {\bibinfo {title} {{The Hagedorn
  temperature of AdS$_5$/CFT$_4$ at finite coupling via the Quantum Spectral
  Curve}},}\ }\href {\doibase 10.1016/j.physletb.2018.09.033} {\bibfield
  {journal} {\bibinfo  {journal} {Phys. Lett. B}\ }\textbf {\bibinfo {volume}
  {786}},\ \bibinfo {pages} {53--58} (\bibinfo {year} {2018}{\natexlab{b}})},\
  \Eprint {http://arxiv.org/abs/1803.04416} {arXiv:1803.04416 [hep-th]}
  \BibitemShut {NoStop}%
\bibitem [{\citenamefont {Harmark}\ and\ \citenamefont
  {Wilhelm}(2022)}]{Harmark:2021qma}%
  \BibitemOpen
  \bibfield  {author} {\bibinfo {author} {\bibfnamefont {Troels}\ \bibnamefont
  {Harmark}}\ and\ \bibinfo {author} {\bibfnamefont {Matthias}\ \bibnamefont
  {Wilhelm}},\ }\bibfield  {title} {\enquote {\bibinfo {title} {{Solving the
  Hagedorn temperature of AdS$_{5}$/CFT$_{4}$ via the Quantum Spectral Curve:
  chemical potentials and deformations}},}\ }\href {\doibase
  10.1007/JHEP07(2022)136} {\bibfield  {journal} {\bibinfo  {journal} {JHEP}\
  }\textbf {\bibinfo {volume} {07}},\ \bibinfo {pages} {136} (\bibinfo {year}
  {2022})},\ \Eprint {http://arxiv.org/abs/2109.09761} {arXiv:2109.09761
  [hep-th]} \BibitemShut {NoStop}%
\bibitem [{\citenamefont {Ekhammar}\ \emph {et~al.}(2023)\citenamefont
  {Ekhammar}, \citenamefont {Minahan},\ and\ \citenamefont
  {Thull}}]{Ekhammar:2023glu}%
  \BibitemOpen
  \bibfield  {author} {\bibinfo {author} {\bibfnamefont {Simon}\ \bibnamefont
  {Ekhammar}}, \bibinfo {author} {\bibfnamefont {Joseph~A.}\ \bibnamefont
  {Minahan}}, \ and\ \bibinfo {author} {\bibfnamefont {Charles}\ \bibnamefont
  {Thull}},\ }\bibfield  {title} {\enquote {\bibinfo {title} {{The asymptotic
  form of the Hagedorn temperature in planar $\mathcal{N}=4$ super
  Yang-Mills}},}\ }\href@noop {} {\  (\bibinfo {year} {2023})},\ \Eprint
  {http://arxiv.org/abs/2306.09883} {arXiv:2306.09883 [hep-th]} \BibitemShut
  {NoStop}%
\bibitem [{\citenamefont {Asplund}\ and\ \citenamefont
  {Berenstein}(2009)}]{Asplund:2008xd}%
  \BibitemOpen
  \bibfield  {author} {\bibinfo {author} {\bibfnamefont {Curtis~T.}\
  \bibnamefont {Asplund}}\ and\ \bibinfo {author} {\bibfnamefont {David}\
  \bibnamefont {Berenstein}},\ }\bibfield  {title} {\enquote {\bibinfo {title}
  {{Small AdS black holes from SYM}},}\ }\href {\doibase
  10.1016/j.physletb.2009.02.043} {\bibfield  {journal} {\bibinfo  {journal}
  {Phys. Lett. B}\ }\textbf {\bibinfo {volume} {673}},\ \bibinfo {pages}
  {264--267} (\bibinfo {year} {2009})},\ \Eprint
  {http://arxiv.org/abs/0809.0712} {arXiv:0809.0712 [hep-th]} \BibitemShut
  {NoStop}%
\bibitem [{\citenamefont {Jokela}\ \emph {et~al.}(2016)\citenamefont {Jokela},
  \citenamefont {P\"onni},\ and\ \citenamefont {Vuorinen}}]{Jokela:2015sza}%
  \BibitemOpen
  \bibfield  {author} {\bibinfo {author} {\bibfnamefont {Niko}\ \bibnamefont
  {Jokela}}, \bibinfo {author} {\bibfnamefont {Arttu}\ \bibnamefont {P\"onni}},
  \ and\ \bibinfo {author} {\bibfnamefont {Aleksi}\ \bibnamefont {Vuorinen}},\
  }\bibfield  {title} {\enquote {\bibinfo {title} {{Small black holes in global
  AdS spacetime}},}\ }\href {\doibase 10.1103/PhysRevD.93.086004} {\bibfield
  {journal} {\bibinfo  {journal} {Phys. Rev. D}\ }\textbf {\bibinfo {volume}
  {93}},\ \bibinfo {pages} {086004} (\bibinfo {year} {2016})},\ \Eprint
  {http://arxiv.org/abs/1508.00859} {arXiv:1508.00859 [hep-th]} \BibitemShut
  {NoStop}%
\bibitem [{\citenamefont {Hanada}\ and\ \citenamefont
  {Maltz}(2017)}]{Hanada:2016pwv}%
  \BibitemOpen
  \bibfield  {author} {\bibinfo {author} {\bibfnamefont {Masanori}\
  \bibnamefont {Hanada}}\ and\ \bibinfo {author} {\bibfnamefont {Jonathan}\
  \bibnamefont {Maltz}},\ }\bibfield  {title} {\enquote {\bibinfo {title} {{A
  proposal of the gauge theory description of the small Schwarzschild black
  hole in AdS$_5\times$S$^5$}},}\ }\href {\doibase 10.1007/JHEP02(2017)012}
  {\bibfield  {journal} {\bibinfo  {journal} {JHEP}\ }\textbf {\bibinfo
  {volume} {02}},\ \bibinfo {pages} {012} (\bibinfo {year} {2017})},\ \Eprint
  {http://arxiv.org/abs/1608.03276} {arXiv:1608.03276 [hep-th]} \BibitemShut
  {NoStop}%
\bibitem [{\citenamefont {Hanada}\ and\ \citenamefont
  {Watanabe}(2023)}]{Hanada:2022wcq}%
  \BibitemOpen
  \bibfield  {author} {\bibinfo {author} {\bibfnamefont {Masanori}\
  \bibnamefont {Hanada}}\ and\ \bibinfo {author} {\bibfnamefont {Hiromasa}\
  \bibnamefont {Watanabe}},\ }\bibfield  {title} {\enquote {\bibinfo {title}
  {{Partial deconfinement: a brief overview}},}\ }\href {\doibase
  10.1140/epjs/s11734-022-00709-0} {\bibfield  {journal} {\bibinfo  {journal}
  {Eur. Phys. J. ST}\ }\textbf {\bibinfo {volume} {232}},\ \bibinfo {pages}
  {333--337} (\bibinfo {year} {2023})},\ \Eprint
  {http://arxiv.org/abs/2210.11216} {arXiv:2210.11216 [hep-th]} \BibitemShut
  {NoStop}%
\bibitem [{\citenamefont {Hanada}\ \emph {et~al.}(2019)\citenamefont {Hanada},
  \citenamefont {Ishiki},\ and\ \citenamefont {Watanabe}}]{Hanada:2018zxn}%
  \BibitemOpen
  \bibfield  {author} {\bibinfo {author} {\bibfnamefont {Masanori}\
  \bibnamefont {Hanada}}, \bibinfo {author} {\bibfnamefont {Goro}\ \bibnamefont
  {Ishiki}}, \ and\ \bibinfo {author} {\bibfnamefont {Hiromasa}\ \bibnamefont
  {Watanabe}},\ }\bibfield  {title} {\enquote {\bibinfo {title} {{Partial
  Deconfinement}},}\ }\href {\doibase 10.1007/JHEP03(2019)145} {\bibfield
  {journal} {\bibinfo  {journal} {JHEP}\ }\textbf {\bibinfo {volume} {03}},\
  \bibinfo {pages} {145} (\bibinfo {year} {2019})},\ \bibinfo {note} {[Erratum:
  JHEP 10, 029 (2019)]},\ \Eprint {http://arxiv.org/abs/1812.05494}
  {arXiv:1812.05494 [hep-th]} \BibitemShut {NoStop}%
\bibitem [{\citenamefont {Gross}\ and\ \citenamefont
  {Witten}(1980)}]{Gross:1980he}%
  \BibitemOpen
  \bibfield  {author} {\bibinfo {author} {\bibfnamefont {D.~J.}\ \bibnamefont
  {Gross}}\ and\ \bibinfo {author} {\bibfnamefont {Edward}\ \bibnamefont
  {Witten}},\ }\bibfield  {title} {\enquote {\bibinfo {title} {{Possible Third
  Order Phase Transition in the Large N Lattice Gauge Theory}},}\ }\href
  {\doibase 10.1103/PhysRevD.21.446} {\bibfield  {journal} {\bibinfo  {journal}
  {Phys. Rev. D}\ }\textbf {\bibinfo {volume} {21}},\ \bibinfo {pages}
  {446--453} (\bibinfo {year} {1980})}\BibitemShut {NoStop}%
\bibitem [{\citenamefont {Wadia}(1980)}]{Wadia:1980cp}%
  \BibitemOpen
  \bibfield  {author} {\bibinfo {author} {\bibfnamefont {Spenta~R.}\
  \bibnamefont {Wadia}},\ }\bibfield  {title} {\enquote {\bibinfo {title} {{$N$
  = Infinity Phase Transition in a Class of Exactly Soluble Model Lattice Gauge
  Theories}},}\ }\href {\doibase 10.1016/0370-2693(80)90353-6} {\bibfield
  {journal} {\bibinfo  {journal} {Phys. Lett. B}\ }\textbf {\bibinfo {volume}
  {93}},\ \bibinfo {pages} {403--410} (\bibinfo {year} {1980})}\BibitemShut
  {NoStop}%
\bibitem [{\citenamefont {Dumitru}\ \emph {et~al.}(2005)\citenamefont
  {Dumitru}, \citenamefont {Lenaghan},\ and\ \citenamefont
  {Pisarski}}]{Dumitru:2004gd}%
  \BibitemOpen
  \bibfield  {author} {\bibinfo {author} {\bibfnamefont {Adrian}\ \bibnamefont
  {Dumitru}}, \bibinfo {author} {\bibfnamefont {Jonathan}\ \bibnamefont
  {Lenaghan}}, \ and\ \bibinfo {author} {\bibfnamefont {Robert~D.}\
  \bibnamefont {Pisarski}},\ }\bibfield  {title} {\enquote {\bibinfo {title}
  {{Deconfinement in matrix models about the Gross-Witten point}},}\ }\href
  {\doibase 10.1103/PhysRevD.71.074004} {\bibfield  {journal} {\bibinfo
  {journal} {Phys. Rev. D}\ }\textbf {\bibinfo {volume} {71}},\ \bibinfo
  {pages} {074004} (\bibinfo {year} {2005})},\ \Eprint
  {http://arxiv.org/abs/hep-ph/0410294} {arXiv:hep-ph/0410294} \BibitemShut
  {NoStop}%
\bibitem [{\citenamefont {Nishimura}\ \emph {et~al.}(2018)\citenamefont
  {Nishimura}, \citenamefont {Pisarski},\ and\ \citenamefont
  {Skokov}}]{Nishimura:2017crr}%
  \BibitemOpen
  \bibfield  {author} {\bibinfo {author} {\bibfnamefont {Hiromichi}\
  \bibnamefont {Nishimura}}, \bibinfo {author} {\bibfnamefont {Robert~D.}\
  \bibnamefont {Pisarski}}, \ and\ \bibinfo {author} {\bibfnamefont
  {Vladimir~V.}\ \bibnamefont {Skokov}},\ }\bibfield  {title} {\enquote
  {\bibinfo {title} {{Finite-temperature phase transitions of third and higher
  order in gauge theories at large $N$}},}\ }\href {\doibase
  10.1103/PhysRevD.97.036014} {\bibfield  {journal} {\bibinfo  {journal} {Phys.
  Rev. D}\ }\textbf {\bibinfo {volume} {97}},\ \bibinfo {pages} {036014}
  (\bibinfo {year} {2018})},\ \Eprint {http://arxiv.org/abs/1712.04465}
  {arXiv:1712.04465 [hep-th]} \BibitemShut {NoStop}%
\bibitem [{\citenamefont {Asano}\ \emph {et~al.}(2020)\citenamefont {Asano},
  \citenamefont {Kov\'a\v{c}ik},\ and\ \citenamefont
  {O'Connor}}]{Asano:2020yry}%
  \BibitemOpen
  \bibfield  {author} {\bibinfo {author} {\bibfnamefont {Yuhma}\ \bibnamefont
  {Asano}}, \bibinfo {author} {\bibfnamefont {Samuel}\ \bibnamefont
  {Kov\'a\v{c}ik}}, \ and\ \bibinfo {author} {\bibfnamefont {Denjoe}\
  \bibnamefont {O'Connor}},\ }\bibfield  {title} {\enquote {\bibinfo {title}
  {{The Confining Transition in the Bosonic BMN Matrix Model}},}\ }\href
  {\doibase 10.1007/JHEP06(2020)174} {\bibfield  {journal} {\bibinfo  {journal}
  {JHEP}\ }\textbf {\bibinfo {volume} {06}},\ \bibinfo {pages} {174} (\bibinfo
  {year} {2020})},\ \Eprint {http://arxiv.org/abs/2001.03749} {arXiv:2001.03749
  [hep-th]} \BibitemShut {NoStop}%
\bibitem [{\citenamefont {Berenstein}(2018)}]{Berenstein:2018lrm}%
  \BibitemOpen
  \bibfield  {author} {\bibinfo {author} {\bibfnamefont {David}\ \bibnamefont
  {Berenstein}},\ }\bibfield  {title} {\enquote {\bibinfo {title} {{Submatrix
  deconfinement and small black holes in AdS}},}\ }\href {\doibase
  10.1007/JHEP09(2018)054} {\bibfield  {journal} {\bibinfo  {journal} {JHEP}\
  }\textbf {\bibinfo {volume} {09}},\ \bibinfo {pages} {054} (\bibinfo {year}
  {2018})},\ \Eprint {http://arxiv.org/abs/1806.05729} {arXiv:1806.05729
  [hep-th]} \BibitemShut {NoStop}%
\bibitem [{\citenamefont {Berenstein}(2019)}]{Berenstein:2018hpl}%
  \BibitemOpen
  \bibfield  {author} {\bibinfo {author} {\bibfnamefont {David}\ \bibnamefont
  {Berenstein}},\ }\bibfield  {title} {\enquote {\bibinfo {title} {{Negative
  specific heat from non-planar interactions and small black holes in
  AdS/CFT}},}\ }\href {\doibase 10.1007/JHEP10(2019)001} {\bibfield  {journal}
  {\bibinfo  {journal} {JHEP}\ }\textbf {\bibinfo {volume} {10}},\ \bibinfo
  {pages} {001} (\bibinfo {year} {2019})},\ \Eprint
  {http://arxiv.org/abs/1810.07267} {arXiv:1810.07267 [hep-th]} \BibitemShut
  {NoStop}%
\bibitem [{\citenamefont {Pak}\ \emph {et~al.}(2019)\citenamefont {Pak},
  \citenamefont {Panova},\ and\ \citenamefont {Yeliussizov}}]{PAK201944}%
  \BibitemOpen
  \bibfield  {author} {\bibinfo {author} {\bibfnamefont {Igor}\ \bibnamefont
  {Pak}}, \bibinfo {author} {\bibfnamefont {Greta}\ \bibnamefont {Panova}}, \
  and\ \bibinfo {author} {\bibfnamefont {Damir}\ \bibnamefont {Yeliussizov}},\
  }\bibfield  {title} {\enquote {\bibinfo {title} {On the largest kronecker and
  littlewood–richardson coefficients},}\ }\href {\doibase
  https://doi.org/10.1016/j.jcta.2019.01.008} {\bibfield  {journal} {\bibinfo
  {journal} {Journal of Combinatorial Theory, Series A}\ }\textbf {\bibinfo
  {volume} {165}},\ \bibinfo {pages} {44--77} (\bibinfo {year}
  {2019})}\BibitemShut {NoStop}%
\bibitem [{\citenamefont {Vershik}\ and\ \citenamefont
  {Kerov}(1985)}]{vershik1985asymptotic}%
  \BibitemOpen
  \bibfield  {author} {\bibinfo {author} {\bibfnamefont {Anatolii~Moiseevich}\
  \bibnamefont {Vershik}}\ and\ \bibinfo {author} {\bibfnamefont
  {Sergei~Vasil'evich}\ \bibnamefont {Kerov}},\ }\bibfield  {title} {\enquote
  {\bibinfo {title} {Asymptotic of the largest and the typical dimensions of
  irreducible representations of a symmetric group},}\ }\href@noop {}
  {\bibfield  {journal} {\bibinfo  {journal} {Funktsional'nyi Analiz i ego
  Prilozheniya}\ }\textbf {\bibinfo {volume} {19}},\ \bibinfo {pages} {25--36}
  (\bibinfo {year} {1985})}\BibitemShut {NoStop}%
\bibitem [{\citenamefont {Logan}\ and\ \citenamefont
  {Shepp}(1977)}]{logan1977variational}%
  \BibitemOpen
  \bibfield  {author} {\bibinfo {author} {\bibfnamefont {Benjamin~F}\
  \bibnamefont {Logan}}\ and\ \bibinfo {author} {\bibfnamefont {Larry~A}\
  \bibnamefont {Shepp}},\ }\bibfield  {title} {\enquote {\bibinfo {title} {A
  variational problem for random young tableaux},}\ }\href@noop {} {\bibfield
  {journal} {\bibinfo  {journal} {Advances in mathematics}\ }\textbf {\bibinfo
  {volume} {26}},\ \bibinfo {pages} {206--222} (\bibinfo {year}
  {1977})}\BibitemShut {NoStop}%
\bibitem [{\citenamefont {Asplund}\ \emph {et~al.}(2013)\citenamefont
  {Asplund}, \citenamefont {Berenstein},\ and\ \citenamefont
  {Dzienkowski}}]{Asplund:2012tg}%
  \BibitemOpen
  \bibfield  {author} {\bibinfo {author} {\bibfnamefont {Curtis~T.}\
  \bibnamefont {Asplund}}, \bibinfo {author} {\bibfnamefont {David}\
  \bibnamefont {Berenstein}}, \ and\ \bibinfo {author} {\bibfnamefont {Eric}\
  \bibnamefont {Dzienkowski}},\ }\bibfield  {title} {\enquote {\bibinfo {title}
  {{Large N classical dynamics of holographic matrix models}},}\ }\href
  {\doibase 10.1103/PhysRevD.87.084044} {\bibfield  {journal} {\bibinfo
  {journal} {Phys. Rev. D}\ }\textbf {\bibinfo {volume} {87}},\ \bibinfo
  {pages} {084044} (\bibinfo {year} {2013})},\ \Eprint
  {http://arxiv.org/abs/1211.3425} {arXiv:1211.3425 [hep-th]} \BibitemShut
  {NoStop}%
\bibitem [{\citenamefont {Corley}\ \emph {et~al.}(2002)\citenamefont {Corley},
  \citenamefont {Jevicki},\ and\ \citenamefont {Ramgoolam}}]{Corley:2001zk}%
  \BibitemOpen
  \bibfield  {author} {\bibinfo {author} {\bibfnamefont {Steve}\ \bibnamefont
  {Corley}}, \bibinfo {author} {\bibfnamefont {Antal}\ \bibnamefont {Jevicki}},
  \ and\ \bibinfo {author} {\bibfnamefont {Sanjaye}\ \bibnamefont
  {Ramgoolam}},\ }\bibfield  {title} {\enquote {\bibinfo {title} {{Exact
  correlators of giant gravitons from dual N=4 SYM theory}},}\ }\href {\doibase
  10.4310/ATMP.2001.v5.n4.a6} {\bibfield  {journal} {\bibinfo  {journal} {Adv.
  Theor. Math. Phys.}\ }\textbf {\bibinfo {volume} {5}},\ \bibinfo {pages}
  {809--839} (\bibinfo {year} {2002})},\ \Eprint
  {http://arxiv.org/abs/hep-th/0111222} {arXiv:hep-th/0111222} \BibitemShut
  {NoStop}%
\bibitem [{\citenamefont {Balasubramanian}\ \emph {et~al.}(2005)\citenamefont
  {Balasubramanian}, \citenamefont {Berenstein}, \citenamefont {Feng},\ and\
  \citenamefont {Huang}}]{Balasubramanian:2004nb}%
  \BibitemOpen
  \bibfield  {author} {\bibinfo {author} {\bibfnamefont {Vijay}\ \bibnamefont
  {Balasubramanian}}, \bibinfo {author} {\bibfnamefont {David}\ \bibnamefont
  {Berenstein}}, \bibinfo {author} {\bibfnamefont {Bo}~\bibnamefont {Feng}}, \
  and\ \bibinfo {author} {\bibfnamefont {Min-xin}\ \bibnamefont {Huang}},\
  }\bibfield  {title} {\enquote {\bibinfo {title} {{D-branes in Yang-Mills
  theory and emergent gauge symmetry}},}\ }\href {\doibase
  10.1088/1126-6708/2005/03/006} {\bibfield  {journal} {\bibinfo  {journal}
  {JHEP}\ }\textbf {\bibinfo {volume} {03}},\ \bibinfo {pages} {006} (\bibinfo
  {year} {2005})},\ \Eprint {http://arxiv.org/abs/hep-th/0411205}
  {arXiv:hep-th/0411205} \BibitemShut {NoStop}%
\bibitem [{\citenamefont {de~Mello~Koch}\ \emph {et~al.}(2007)\citenamefont
  {de~Mello~Koch}, \citenamefont {Smolic},\ and\ \citenamefont
  {Smolic}}]{deMelloKoch:2007rqf}%
  \BibitemOpen
  \bibfield  {author} {\bibinfo {author} {\bibfnamefont {Robert}\ \bibnamefont
  {de~Mello~Koch}}, \bibinfo {author} {\bibfnamefont {Jelena}\ \bibnamefont
  {Smolic}}, \ and\ \bibinfo {author} {\bibfnamefont {Milena}\ \bibnamefont
  {Smolic}},\ }\bibfield  {title} {\enquote {\bibinfo {title} {{Giant Gravitons
  - with Strings Attached (I)}},}\ }\href {\doibase
  10.1088/1126-6708/2007/06/074} {\bibfield  {journal} {\bibinfo  {journal}
  {JHEP}\ }\textbf {\bibinfo {volume} {06}},\ \bibinfo {pages} {074} (\bibinfo
  {year} {2007})},\ \Eprint {http://arxiv.org/abs/hep-th/0701066}
  {arXiv:hep-th/0701066} \BibitemShut {NoStop}%
\bibitem [{\citenamefont {Brown}\ \emph {et~al.}(2008)\citenamefont {Brown},
  \citenamefont {Heslop},\ and\ \citenamefont {Ramgoolam}}]{Brown:2007xh}%
  \BibitemOpen
  \bibfield  {author} {\bibinfo {author} {\bibfnamefont {Thomas~William}\
  \bibnamefont {Brown}}, \bibinfo {author} {\bibfnamefont {P.~J.}\ \bibnamefont
  {Heslop}}, \ and\ \bibinfo {author} {\bibfnamefont {S.}~\bibnamefont
  {Ramgoolam}},\ }\bibfield  {title} {\enquote {\bibinfo {title} {{Diagonal
  multi-matrix correlators and BPS operators in N=4 SYM}},}\ }\href {\doibase
  10.1088/1126-6708/2008/02/030} {\bibfield  {journal} {\bibinfo  {journal}
  {JHEP}\ }\textbf {\bibinfo {volume} {02}},\ \bibinfo {pages} {030} (\bibinfo
  {year} {2008})},\ \Eprint {http://arxiv.org/abs/0711.0176} {arXiv:0711.0176
  [hep-th]} \BibitemShut {NoStop}%
\bibitem [{\citenamefont {Bhattacharyya}\ \emph
  {et~al.}(2008{\natexlab{a}})\citenamefont {Bhattacharyya}, \citenamefont
  {Collins},\ and\ \citenamefont {de~Mello~Koch}}]{Bhattacharyya:2008rb}%
  \BibitemOpen
  \bibfield  {author} {\bibinfo {author} {\bibfnamefont {Rajsekhar}\
  \bibnamefont {Bhattacharyya}}, \bibinfo {author} {\bibfnamefont {Storm}\
  \bibnamefont {Collins}}, \ and\ \bibinfo {author} {\bibfnamefont {Robert}\
  \bibnamefont {de~Mello~Koch}},\ }\bibfield  {title} {\enquote {\bibinfo
  {title} {{Exact Multi-Matrix Correlators}},}\ }\href {\doibase
  10.1088/1126-6708/2008/03/044} {\bibfield  {journal} {\bibinfo  {journal}
  {JHEP}\ }\textbf {\bibinfo {volume} {03}},\ \bibinfo {pages} {044} (\bibinfo
  {year} {2008}{\natexlab{a}})},\ \Eprint {http://arxiv.org/abs/0801.2061}
  {arXiv:0801.2061 [hep-th]} \BibitemShut {NoStop}%
\bibitem [{\citenamefont {Bhattacharyya}\ \emph
  {et~al.}(2008{\natexlab{b}})\citenamefont {Bhattacharyya}, \citenamefont
  {de~Mello~Koch},\ and\ \citenamefont {Stephanou}}]{Bhattacharyya:2008xy}%
  \BibitemOpen
  \bibfield  {author} {\bibinfo {author} {\bibfnamefont {Rajsekhar}\
  \bibnamefont {Bhattacharyya}}, \bibinfo {author} {\bibfnamefont {Robert}\
  \bibnamefont {de~Mello~Koch}}, \ and\ \bibinfo {author} {\bibfnamefont
  {Michael}\ \bibnamefont {Stephanou}},\ }\bibfield  {title} {\enquote
  {\bibinfo {title} {{Exact Multi-Restricted Schur Polynomial Correlators}},}\
  }\href {\doibase 10.1088/1126-6708/2008/06/101} {\bibfield  {journal}
  {\bibinfo  {journal} {JHEP}\ }\textbf {\bibinfo {volume} {06}},\ \bibinfo
  {pages} {101} (\bibinfo {year} {2008}{\natexlab{b}})},\ \Eprint
  {http://arxiv.org/abs/0805.3025} {arXiv:0805.3025 [hep-th]} \BibitemShut
  {NoStop}%
\bibitem [{\citenamefont {de~Mello~Koch}\ and\ \citenamefont
  {Ramgoolam}(2012)}]{deMelloKoch:2012ck}%
  \BibitemOpen
  \bibfield  {author} {\bibinfo {author} {\bibfnamefont {Robert}\ \bibnamefont
  {de~Mello~Koch}}\ and\ \bibinfo {author} {\bibfnamefont {Sanjaye}\
  \bibnamefont {Ramgoolam}},\ }\bibfield  {title} {\enquote {\bibinfo {title}
  {{A double coset ansatz for integrability in AdS/CFT}},}\ }\href {\doibase
  10.1007/JHEP06(2012)083} {\bibfield  {journal} {\bibinfo  {journal} {JHEP}\
  }\textbf {\bibinfo {volume} {06}},\ \bibinfo {pages} {083} (\bibinfo {year}
  {2012})},\ \Eprint {http://arxiv.org/abs/1204.2153} {arXiv:1204.2153
  [hep-th]} \BibitemShut {NoStop}%
\bibitem [{\citenamefont {Pasukonis}\ and\ \citenamefont
  {Ramgoolam}(2013)}]{Pasukonis:2013ts}%
  \BibitemOpen
  \bibfield  {author} {\bibinfo {author} {\bibfnamefont {Jurgis}\ \bibnamefont
  {Pasukonis}}\ and\ \bibinfo {author} {\bibfnamefont {Sanjaye}\ \bibnamefont
  {Ramgoolam}},\ }\bibfield  {title} {\enquote {\bibinfo {title} {{Quivers as
  Calculators: Counting, Correlators and Riemann Surfaces}},}\ }\href {\doibase
  10.1007/JHEP04(2013)094} {\bibfield  {journal} {\bibinfo  {journal} {JHEP}\
  }\textbf {\bibinfo {volume} {04}},\ \bibinfo {pages} {094} (\bibinfo {year}
  {2013})},\ \Eprint {http://arxiv.org/abs/1301.1980} {arXiv:1301.1980
  [hep-th]} \BibitemShut {NoStop}%
\bibitem [{\citenamefont {Berenstein}(2015)}]{Berenstein:2015ooa}%
  \BibitemOpen
  \bibfield  {author} {\bibinfo {author} {\bibfnamefont {David}\ \bibnamefont
  {Berenstein}},\ }\bibfield  {title} {\enquote {\bibinfo {title} {{Extremal
  chiral ring states in the AdS/CFT correspondence are described by free
  fermions for a generalized oscillator algebra}},}\ }\href {\doibase
  10.1103/PhysRevD.92.046006} {\bibfield  {journal} {\bibinfo  {journal} {Phys.
  Rev. D}\ }\textbf {\bibinfo {volume} {92}},\ \bibinfo {pages} {046006}
  (\bibinfo {year} {2015})},\ \Eprint {http://arxiv.org/abs/1504.05389}
  {arXiv:1504.05389 [hep-th]} \BibitemShut {NoStop}%
\bibitem [{\citenamefont {Berenstein}(2005)}]{Berenstein:2004hw}%
  \BibitemOpen
  \bibfield  {author} {\bibinfo {author} {\bibfnamefont {David}\ \bibnamefont
  {Berenstein}},\ }\bibfield  {title} {\enquote {\bibinfo {title} {{A Matrix
  model for a quantum Hall droplet with manifest particle-hole symmetry}},}\
  }\href {\doibase 10.1103/PhysRevD.71.085001} {\bibfield  {journal} {\bibinfo
  {journal} {Phys. Rev. D}\ }\textbf {\bibinfo {volume} {71}},\ \bibinfo
  {pages} {085001} (\bibinfo {year} {2005})},\ \Eprint
  {http://arxiv.org/abs/hep-th/0409115} {arXiv:hep-th/0409115} \BibitemShut
  {NoStop}%
\bibitem [{\citenamefont {Berenstein}\ and\ \citenamefont
  {de~Mello~Koch}(2019)}]{Berenstein:2019esh}%
  \BibitemOpen
  \bibfield  {author} {\bibinfo {author} {\bibfnamefont {David}\ \bibnamefont
  {Berenstein}}\ and\ \bibinfo {author} {\bibfnamefont {Robert}\ \bibnamefont
  {de~Mello~Koch}},\ }\bibfield  {title} {\enquote {\bibinfo {title} {{Gauged
  fermionic matrix quantum mechanics}},}\ }\href {\doibase
  10.1007/JHEP03(2019)185} {\bibfield  {journal} {\bibinfo  {journal} {JHEP}\
  }\textbf {\bibinfo {volume} {03}},\ \bibinfo {pages} {185} (\bibinfo {year}
  {2019})},\ \Eprint {http://arxiv.org/abs/1903.01628} {arXiv:1903.01628
  [hep-th]} \BibitemShut {NoStop}%
\bibitem [{\citenamefont {Collins}(2009)}]{Collins:2008gc}%
  \BibitemOpen
  \bibfield  {author} {\bibinfo {author} {\bibfnamefont {Storm}\ \bibnamefont
  {Collins}},\ }\bibfield  {title} {\enquote {\bibinfo {title} {{Restricted
  Schur Polynomials and Finite N Counting}},}\ }\href {\doibase
  10.1103/PhysRevD.79.026002} {\bibfield  {journal} {\bibinfo  {journal} {Phys.
  Rev. D}\ }\textbf {\bibinfo {volume} {79}},\ \bibinfo {pages} {026002}
  (\bibinfo {year} {2009})},\ \Eprint {http://arxiv.org/abs/0810.4217}
  {arXiv:0810.4217 [hep-th]} \BibitemShut {NoStop}%
\bibitem [{\citenamefont {Mattioli}\ and\ \citenamefont
  {Ramgoolam}(2016)}]{Mattioli:2016eyp}%
  \BibitemOpen
  \bibfield  {author} {\bibinfo {author} {\bibfnamefont {Paolo}\ \bibnamefont
  {Mattioli}}\ and\ \bibinfo {author} {\bibfnamefont {Sanjaye}\ \bibnamefont
  {Ramgoolam}},\ }\bibfield  {title} {\enquote {\bibinfo {title} {{Permutation
  Centralizer Algebras and Multi-Matrix Invariants}},}\ }\href {\doibase
  10.1103/PhysRevD.93.065040} {\bibfield  {journal} {\bibinfo  {journal} {Phys.
  Rev. D}\ }\textbf {\bibinfo {volume} {93}},\ \bibinfo {pages} {065040}
  (\bibinfo {year} {2016})},\ \Eprint {http://arxiv.org/abs/1601.06086}
  {arXiv:1601.06086 [hep-th]} \BibitemShut {NoStop}%
\bibitem [{\citenamefont {de~Mello~Koch}\ \emph {et~al.}(2013)\citenamefont
  {de~Mello~Koch}, \citenamefont {Diaz},\ and\ \citenamefont
  {Nokwara}}]{deMelloKoch:2012sie}%
  \BibitemOpen
  \bibfield  {author} {\bibinfo {author} {\bibfnamefont {Robert}\ \bibnamefont
  {de~Mello~Koch}}, \bibinfo {author} {\bibfnamefont {Pablo}\ \bibnamefont
  {Diaz}}, \ and\ \bibinfo {author} {\bibfnamefont {Nkululeko}\ \bibnamefont
  {Nokwara}},\ }\bibfield  {title} {\enquote {\bibinfo {title} {{Restricted
  Schur Polynomials for Fermions and integrability in the su(2|3) sector}},}\
  }\href {\doibase 10.1007/JHEP03(2013)173} {\bibfield  {journal} {\bibinfo
  {journal} {JHEP}\ }\textbf {\bibinfo {volume} {03}},\ \bibinfo {pages} {173}
  (\bibinfo {year} {2013})},\ \Eprint {http://arxiv.org/abs/1212.5935}
  {arXiv:1212.5935 [hep-th]} \BibitemShut {NoStop}%
\bibitem [{\citenamefont {Ramgoolam}(2016)}]{Ramgoolam:2016ciq}%
  \BibitemOpen
  \bibfield  {author} {\bibinfo {author} {\bibfnamefont {Sanjaye}\ \bibnamefont
  {Ramgoolam}},\ }\bibfield  {title} {\enquote {\bibinfo {title} {{Permutations
  and the combinatorics of gauge invariants for general N}},}\ }\href {\doibase
  10.22323/1.263.0107} {\bibfield  {journal} {\bibinfo  {journal} {PoS}\
  }\textbf {\bibinfo {volume} {CORFU2015}},\ \bibinfo {pages} {107} (\bibinfo
  {year} {2016})},\ \Eprint {http://arxiv.org/abs/1605.00843} {arXiv:1605.00843
  [hep-th]} \BibitemShut {NoStop}%
\bibitem [{\citenamefont {Harris}\ and\ \citenamefont
  {Willenbring}(2014)}]{harris2014sums}%
  \BibitemOpen
  \bibfield  {author} {\bibinfo {author} {\bibfnamefont {Pamela~E}\
  \bibnamefont {Harris}}\ and\ \bibinfo {author} {\bibfnamefont {Jeb~F}\
  \bibnamefont {Willenbring}},\ }\bibfield  {title} {\enquote {\bibinfo {title}
  {Sums of squares of littlewood--richardson coefficients and gl n-harmonic
  polynomials},}\ }\href@noop {} {\bibfield  {journal} {\bibinfo  {journal}
  {Symmetry: Representation Theory and Its Applications: In Honor of Nolan R.
  Wallach}\ ,\ \bibinfo {pages} {305--326}} (\bibinfo {year}
  {2014})}\BibitemShut {NoStop}%
\bibitem [{\citenamefont {O'Connor}(2022)}]{O'connor}%
  \BibitemOpen
  \bibfield  {author} {\bibinfo {author} {\bibfnamefont {Denjoe}\ \bibnamefont
  {O'Connor}},\ }\href@noop {} {\emph {\bibinfo {title} {The Hagedorn
  Transition in the Bosonic BFSS Model Revisited}}}\ (\bibinfo  {publisher}
  {Talk given at the {"Non-perturbative and numerical approaches to quantum
  gravity, string theory and Holography"} workshop, ICTS},\ \bibinfo {year}
  {2022})\BibitemShut {NoStop}%
\bibitem [{\citenamefont {Pisarski}\ and\ \citenamefont
  {Skokov}(2012)}]{Pisarski:2012bj}%
  \BibitemOpen
  \bibfield  {author} {\bibinfo {author} {\bibfnamefont {Robert~D.}\
  \bibnamefont {Pisarski}}\ and\ \bibinfo {author} {\bibfnamefont
  {Vladimir~V.}\ \bibnamefont {Skokov}},\ }\bibfield  {title} {\enquote
  {\bibinfo {title} {{Gross-Witten-Wadia transition in a matrix model of
  deconfinement}},}\ }\href {\doibase 10.1103/PhysRevD.86.081701} {\bibfield
  {journal} {\bibinfo  {journal} {Phys. Rev. D}\ }\textbf {\bibinfo {volume}
  {86}},\ \bibinfo {pages} {081701} (\bibinfo {year} {2012})},\ \Eprint
  {http://arxiv.org/abs/1206.1329} {arXiv:1206.1329 [hep-th]} \BibitemShut
  {NoStop}%
\bibitem [{\citenamefont {Narayanan}(2005)}]{narayanan2005computation}%
  \BibitemOpen
  \bibfield  {author} {\bibinfo {author} {\bibfnamefont {Hariharan}\
  \bibnamefont {Narayanan}},\ }\href@noop {} {\enquote {\bibinfo {title} {The
  computation of kostka numbers and littlewood-richardson coefficients is \#\!
  p-complete},}\ } (\bibinfo {year} {2005}),\ \Eprint
  {http://arxiv.org/abs/math/0501176} {arXiv:math/0501176 [math.CO]}
  \BibitemShut {NoStop}%
\bibitem [{\citenamefont {Berenstein}\ and\ \citenamefont
  {Miller}(2017)}]{Berenstein:2017abm}%
  \BibitemOpen
  \bibfield  {author} {\bibinfo {author} {\bibfnamefont {David}\ \bibnamefont
  {Berenstein}}\ and\ \bibinfo {author} {\bibfnamefont {Alexandra}\
  \bibnamefont {Miller}},\ }\bibfield  {title} {\enquote {\bibinfo {title}
  {{Superposition induced topology changes in quantum gravity}},}\ }\href
  {\doibase 10.1007/JHEP11(2017)121} {\bibfield  {journal} {\bibinfo  {journal}
  {JHEP}\ }\textbf {\bibinfo {volume} {11}},\ \bibinfo {pages} {121} (\bibinfo
  {year} {2017})},\ \Eprint {http://arxiv.org/abs/1702.03011} {arXiv:1702.03011
  [hep-th]} \BibitemShut {NoStop}%
\bibitem [{\citenamefont {Berenstein}\ and\ \citenamefont
  {Miller}(2018)}]{Berenstein:2017rrx}%
  \BibitemOpen
  \bibfield  {author} {\bibinfo {author} {\bibfnamefont {David}\ \bibnamefont
  {Berenstein}}\ and\ \bibinfo {author} {\bibfnamefont {Alexandra}\
  \bibnamefont {Miller}},\ }\bibfield  {title} {\enquote {\bibinfo {title}
  {{Code subspaces for LLM geometries}},}\ }\href {\doibase
  10.1088/1361-6382/aaa623} {\bibfield  {journal} {\bibinfo  {journal} {Class.
  Quant. Grav.}\ }\textbf {\bibinfo {volume} {35}},\ \bibinfo {pages} {065003}
  (\bibinfo {year} {2018})},\ \Eprint {http://arxiv.org/abs/1708.00035}
  {arXiv:1708.00035 [hep-th]} \BibitemShut {NoStop}%
\bibitem [{\citenamefont {Balasubramanian}\ \emph {et~al.}(2019)\citenamefont
  {Balasubramanian}, \citenamefont {Berenstein}, \citenamefont {Lewkowycz},
  \citenamefont {Miller}, \citenamefont {Parrikar},\ and\ \citenamefont
  {Rabideau}}]{Balasubramanian:2018yjq}%
  \BibitemOpen
  \bibfield  {author} {\bibinfo {author} {\bibfnamefont {Vijay}\ \bibnamefont
  {Balasubramanian}}, \bibinfo {author} {\bibfnamefont {David}\ \bibnamefont
  {Berenstein}}, \bibinfo {author} {\bibfnamefont {Aitor}\ \bibnamefont
  {Lewkowycz}}, \bibinfo {author} {\bibfnamefont {Alexandra}\ \bibnamefont
  {Miller}}, \bibinfo {author} {\bibfnamefont {Onkar}\ \bibnamefont
  {Parrikar}}, \ and\ \bibinfo {author} {\bibfnamefont {Charles}\ \bibnamefont
  {Rabideau}},\ }\bibfield  {title} {\enquote {\bibinfo {title} {{Emergent
  classical spacetime from microstates of an incipient black hole}},}\ }\href
  {\doibase 10.1007/JHEP01(2019)197} {\bibfield  {journal} {\bibinfo  {journal}
  {JHEP}\ }\textbf {\bibinfo {volume} {01}},\ \bibinfo {pages} {197} (\bibinfo
  {year} {2019})},\ \Eprint {http://arxiv.org/abs/1810.13440} {arXiv:1810.13440
  [hep-th]} \BibitemShut {NoStop}%
\end{thebibliography}%

\end{document}